# The size-distribution of Scattered Disk TNOs from that of JFCs between 0.2 and 15 km effective radius.

Michael Belton

*Belton Space Exploration Initiatives, 430 S. Randolph Way, Tucson, AZ 85716.
Emeritus astronomer, National Optical Astronomy Observatories, Tucson, Arizona*

Pages      47

Figures    12

Tables      2

Corresponding author:

Michael J.S. Belton

Belton Space Exploration Initiatives, LLC

430 S Randolph Way

Tucson, AZ 85716

Phone:	520-795-6286

Cell:	520-275-0066

Fax:	520-795-6286

Email:	mbelton@dakotacom.net

# ABSTRACT

We investigate the differential size-frequency distribution (SFD) of Jupiter Family comets (JFCs) in order to determine whether they are primordial accreted objects or collisional fragments as suggested by current models of the evolution of Trans-Neptunian Objects (TNOs). We develop a list of effective radii and their uncertainties for 161 active JFCs from published sources and compute the observed differential size-frequency distribution using a Probability Index technique. The radii range from 0.2 to 15.4 km and average 1.9 km. The peak of the distribution is near 1.0 km. This is then corrected for the effects of observational selection using a model published earlier by Meech *et al.* (Icarus 170, 463-491, 2004). We estimate that the total number of active JFCs between 0.2 and 15.4 km is approximately 2300 indicating that our current sample of the of active JFC population is far from complete. The active JFC size-frequency distribution, over the range from 0.6 to 10 km where it is best defined, is found to be closer to an exponential distribution in character than a power-law. We then develop a statistical model, based on the assumption of a steady state, for converting the distribution of active JFCs to the SFD of the source population among the TNOs. The model includes the effects of devolatization (that produces a large sub-class of defunct nuclei) and surficial mass-loss. Comparison with available TNO observations shows that to simultaneously attain continuity with the data on objects in the hot TNO population (Fuentes *et al.* (Astrophys.J 722,1290-1304; 2010), satisfy constraints on the number of TNOs set by the occultation detections of Schlichting *et al.* (Ap.J. 761:150; 2012), and to remain within upper limits set by the Taiwanese-American Occultation Survey (TAOS; Zhang *et al*, Astron. J. 146, Id 14, 10pp) the total JFC population must contain a large fraction of small defunct nuclei. The effective power-law index of the inferred TNO differential SFD between 1 and 10 km is -4.5 +/- 0.5 indicating a population in this range that is not in fully relaxed collisional equilibrium. We conclude that the cometary nuclei so far visited by spacecraft and many JFCs are primordial accreted objects relatively unaffected by collisional evolution. We find a turndown in the slope of the predicted TNO cumulative distribution near 1 km radius rather than near 10 km that is seen in many TNO evolutionary calculations. This may or may not represent the onset of a collisional cascade.

1. **Introduction and basic assumptions**

Are small objects (0.1 to 10 km in radius) in the scattered disk TNO (Trans-Neptunian Object) population, which are the putative source population of comet nuclei observed today as Jupiter family comets (JFCs), representative samples of primitive accreted objects much as first outlined by Weidenschilling (1997), or are they more representative of collisional fragments as suggested by the calculations of Davis and Farinella (1997), or even collisionally processed aggregates (primitive rubble piles) as discussed by Weisman *et al.* (2004)?

Farinella and Davis (1996) addressed a version of this question and their modeling indicates that TNOs in the relevant size range would be part of a cascade of collisional fragments that would display a power-law size-frequency distribution (SFD) with a differential index near -3.5. Recent modeling (see Kenyon *et al.* (2008) for a review) continues to support this view as well as giving a more detailed idea of what the shape of the cumulative SFD in this size range might look like. The most recent modeling (Schlichting *et al*, 2013) continues to show how important the effects of collisions can be on the evolving SFD. In parallel with this modeling there has been a concerted effort by several groups of Earth-based observers to determine the SFD of JFCs and measure the slope of its cumulative SFD (most recently Fernández *et al*, 2013; Weiler *et al.,* 2011; Snodgrass *et al.,* 2011; Tancredi *et al.,* 2006; and Meech *et al.*, 2004). Their results for the cumulative power-law index, -2.01 ± 0.21; -1.9 ± 0.2; -1.92 ± 0.20 ( $r > 1.25$ km), -2.7 ± 0.3 (obtained from the absolute nuclear magnitude range of ~14 to 16 mag), and -1.45 ± 0.05 ( $1 < r < 10$ km) respectively, are sensitive to the size range employed and, except for the result of Tancredi *et al.* (2006), are all some way away from the cumulative size-distribution slope of -2.5 expected for a collisional cascade in equilibrium (Dohnanyi, 1969), but not necessarily the predicted cumulative slope of -2.04 in a steady-state population of nuclei with no material strength by O'Brian and Greenberg (2003). It is of interest to note that the three most recent studies give effectively identical results while the outlier at -2.7 is probably due to the short baseline in effective radius (2.1 to 5.3 km).

Now, however, we have certain aspects of spacecraft observations of JFCs that question the collisional scenario outlined above. Brownlee *et al.* (2004) and Weaver (2004) were the first to suggest, on the basis of high-resolution Stardust Mission images, that the rounded shape of the nucleus of 81P/Wild2 did not seem consistent with a collisional fragment and that it was probably not a rubble pile. They considered that their results presented a "challenge" to the collisionally processed aggregate paradigm. Subsequent discovery of deep-seated, global-scale, layers on 9P/Tempel 1 in the data from the Deep Impact and Stardust-Next missions (Thomas *et al.,* 2007, Thomas *et al.*, 2013a, b) and their interpretation as primordial layers by Belton *et al.* (2007) also appears to be in conflict with "models of fragmentation and rubble pile building." Similarly, there is the lack of any convincing or, at least, non-controversial, evidence of wide-spread impact cratering on the surfaces of the observed nuclei as predicted by Durda and Stern (2000). While the globally distributed pits seen on 81P/Wild 2 (Brownlee *et al.,* 2004) have been interpreted as impact craters (Basilevsky and Keller, 2007), they are, in our opinion, more likely to be sublimational or erosional pits caused by internal or outburst activity (Belton*,* 2010). Similarly, the widespread distribution of pits seen on 9P/Tempel 1 are more reasonably explained by comet outburst activity than by impact (Belton *et al.,* 2013).

It therefore appears that there may be a serious conflict between expectations for comet nuclei based on modeling of collisional evolution in the Kuiper belt and what high-resolution geophysical studies of individual nuclei tell us. A resolution may possibly come from two directions. First, many of the Kuiper Belt evolution calculations depend on a fragmentation model that was developed for understanding the evolution of rocky asteroids that suffer hypersonic collisions. As far as we are aware a competent model of *comet fragmentation at sub-sonic speeds* does not yet exist and it may be that the 'rocky' model of fragmentation is not appropriate for TNOs. The collisional cascade SFD of real comet nuclei may have a different character, *i.e.,* shape and slope, from that which is currently predicted. Secondly, there are other uncertainties in this kind of modeling that involve the coupling of dynamical and collisional evolution. These are, in our opinion, well exposed in the work of Charnoz and Morbidelli (2003, 2007) and could lead to a different assessment of the effects of collisional evolution. On the other hand, it surely can be argued that since the data from spacecraft missions is relatively new, current results and interpretations are not assured. So there is ample opportunity for new interpretations to be devised for what is observed.

Our approach to resolving these problems is two-fold: First, establish the differential SFD of scattered-disk TNOs (SDOs) in a size (effective radius) range relevant to the mission-observed comet nuclei, *i.e.,* JFCs, and compare with the results of models of collisional erosion. We do this by an extrapolation from the observed SFD of JFCs taking into account the effects of observational selection, devolatization, and mass-loss. Second, establish a model of comet impact fragmentation for low to moderate relative speeds that is based on the many astronomical observations of the phenomenon, and, for higher speeds, the results of the hypersonic Deep Impact experiment (Schultz *et al.,* 2007; Richardson *et al.,* 2007). The first Item is the focus of this paper; the second investigation is reserved for a future paper that is currently in preparation.

To proceed it is necessary to make some basic assumptions: First, we will accept the result that a dynamical connection currently exists between the SDOs and JFCs as originally worked out by Levison and Duncan (1997) and Duncan *et al.* (2004) and adopt the timescales for the various dynamical processes that were established in their work. Secondly, we will assume that both the SDOs and JFC size frequency distributions (averaged over a timescale of ~ $10^4$ y [see below]) are currently in a *steady-state*, specifically that they are independent of time and that the dynamical processes involved in the transfer of an SDO to a JFC nucleus are independent of size. There is no observational evidence to support the steady-state assumption that we are aware of, however, the evolutionary calculations of Volk (2013) and Brasser and Morbidelli (2013) indicate that, while systematic changes in the scattered disk are expected to occur, their effects are quite small over timescales of $10^4 – 10^8$ y (see Volk (2013) for a detailed discussion of the steady state hypothesis and the possible effect of an episodic fluctuation). Finally, we will also assume that *(i)* the density distribution in the interior of comet nuclei is homogeneous (Belton *et al.*, 1991): *(ii)* that the individual shapes of comet nuclei do not affect the calculations, *(iii)* that their size can be represented by an *effective radius*, $r$, and, *(iv)* that the effects of mass-loss on size while a comet nucleus is in its Centaur stage are negligible (for a review of activity in these objects see Jewitt, 2009).

In section 2 we establish the statistical relationship between the JFC and SDO differential SFDs. In section 3 we provide background on the observed SDOs and JFC size frequency distribution and establish the observed JFC size frequency distribution that we will use in this work. This section also discusses the relevant orbital properties of the JFC sample that we use. In section 4 we adjust the observed SFD for observational selection effects based on a model suggested by Meech *et al.* (2004). In section 5 we compute the SFD for scattered disk TNOs by simultaneously constraining its parameters with three independent observations of the hot TNO population. Finally, in section 6, we provide a discussion of our results in terms of the question posed above.

## 2. The relationship between the JFC and SDs size-frequency distributions.

To be clear, we first define what is included in the sub-set of comets that we call JFCs. Following Duncan *et al.* (2004) we group comets according to their Tisserand parameter, $T$ (defined with respect to Jupiter), and orbital inclination, $i$ and define JFCs as those comets that lie between the following limits:

$$2.4 \leq T \leq 3.1 \quad \text{and} \quad 0 \leq i \leq 31° \tag{1}$$

The usual definition of JFCs is $2 < T < 3$ (Duncan *et al.*, 2004) and our adjustment of these limits is to include comets like 2P/Encke and a few others that have a Tisserand parameter slightly greater than 3 and to definitely exclude Oort type comets. Our chosen range ensures that the comets we employ likely originated as SDOs and have subsequently been introduced into their current orbits by dynamical interaction with Jupiter (see the review by Duncan *et al.,* 2004). As discussed in the following section, the limits in Eq. (1) also reflect the population of JFCs for which a size determination is available and there are 161 in our list (Table 1). In what follows we also include the possibility of a sub-class of defunct nuclei in the JFC population. "Defunct" is a term used by Jewitt (2004) to describe both dead, *i.e.,* exhausted of volatiles, and dormant comets, *i.e.,* comets that have developed thick mantles that prevent further sublimation of volatiles. There is substantial observational evidence for the existence of such a sub-class for, in a study of a different but related population, the near-Earth objects, Fernández *et al.* (2002) found that up to 20% may be "dead", or dormant, nuclei originally from the JFC population. Except in the case of the Near Earth Object (NEO) population and the most recent MPC catalog (as of October 2013), only a few "asteroidal" objects on cometary orbits have so far been found (Licandro et al., 2008; Weissman *et al.,* 2002; Jewitt, 2004). Licandro et al. (2008) developed a list of 41 asteroids in cometary orbits (ACOs) of which 84% showed featureless spectra that strongly suggest a cometary origin. 25 of these ACOs meet our criteria (Eq. 1) for JFC membership. Further discussion of ACOs and their possible relationship to dormant comets is reserved for section 6.

Let $J(r)$, $D(r)$ and $S(r)$ be the differential SFDs of active and defunct nuclei in the JFC population and in the JFC *source* population respectively. $S(r)$ is presumably a sub-set of the Centaur population. $J(r)dr, D(r)dr$ and $S(r)dr$ represent the number of objects in the three populations with sizes (effective radii) between $r$ and $r + dr$. We have already assumed that the dynamical interactions that inject new JFCs and that cause their loss from the existing JFC population are independent of $r$, therefore, providing that JFCs nuclei do not suffer changes in their size during their tenure in the JFC population, in a *steady state* we would have equality between the injection rate and loss rate from $J(r) + D(r)$ when averaged over many orbital periods, *i.e.*:

$$\alpha S(r) = \left(\frac{1}{\tau}\right)(J(r) + D(r)) \qquad \text{[zero mass-loss case]} \qquad (2)$$

where the coefficient $\alpha$ is the fraction of the source population injected per unit time as new JFCs, and $\tau$ is the dynamical lifetime of the entire JFC population. Note that there is no term in Eq.2 that describes the change from the active state to the defunct state since the number of objects is conserved in that process. It is, however, plausible to assume that the source population only injects nuclei capable of becoming active in their early tenure in the JFC population, *i.e,* the source population does not inject defunct nuclei. If this is true, we can balance the dynamical loss of defunct nuclei with their gain from devolatized active nuclei and write for the defunct sub-class in the steady state:

$$\beta(r)J(r) - (1/\tau)D(r) = 0 \qquad (3)$$

or

$$\frac{D(r)}{J(r)} = \tau \beta(r) \qquad (4)$$

where $\beta(r)$ describes the rate of transformation from the active to the defunct state.

Jewitt (2004; his Eq. 20) previously provided a form of this equation in which he identifies $\beta(r)$ as $1/\tau_{dv}$ where $\tau_{dv}$ is a devolatization timescale that is proportional to r. To proceed, we will adopt Jewitt's identification here writing $\beta(r) = \frac{\beta_0}{r}$ while recognizing that the true situation may be substantially more complex and need revision in the future. This allows Eq. 2 to be rewritten as:

$$\alpha S(r) = \left(\frac{1}{\tau}\right)\left(1 + \frac{\tau \beta_0}{r}\right) J(r) \qquad (5)$$

[We note that Jewitt's Eq.20 does not explicitly recognize the entire r-dependence expressed in our Eq. 4 but equates the ratio of the *total* numbers of dead to active comets to $\tau \beta(r)$ and suggests that this ratio is between 2 and 6.7 (the latter number from Levison and Duncan, 1997). If we follow Snodgrass *et al.* (2011) and write the cumulative SFD of JFCs as $r^{-1.92}$ then, integrating over the range 0.1 < r < 10 km, the ratio is approximately 14, with the preponderance of dead comets being at the small end.]

Eq. 5, however, cannot be true if there is an ongoing transformation of larger objects into smaller ones*, i.e.,* if mass-loss is occurring. We account for this with an additional term that describes the difference between the rates at which nuclei with radius $r$ are created and removed. For a steady state:

$$S(r) = \left(\frac{1}{\alpha \tau}\right)\left[\left(1 + \frac{\tau \beta_0}{r}\right) J(r) + \tau \dot{r} \frac{dJ(r)}{dr}\right] \qquad \text{[mass-loss case]} \qquad (6)$$

where $\dot{r}$ is the rate of change of effective radius $r$ due to mass-loss averaged over the entire distribution of JFCs and $\beta_0$ controls the relative population of defunct nuclei. Below, we shall show that for the dominant mass-loss mechanism, $\dot{r}$ can be taken as independent of $r$. We have not included in Eq. 6 a specific term that describes the production of secondary (or tertiary) fragments following the splitting mass-loss process. The reason for this is that such fragments are generally < 100m in effective radius and are short lived (Boehnhardt, 2004). Such fragments fall outside of the range of radii that we are considering and, in any event, because of their short life can be thought of part of the defunct population and can be accounted for through the devolatization term. Weissman and Lowry (2003) (see also Lowry *et al.* 2008) have investigated the effects of $H_2O$ sublimation mass-loss on the shape of the cumulative form of J(r) and find that the cumulative power-law slope might change by about -0.1 in the power-law index*, i.e.,* a small effect.

2.1 *Some considerations on mass-loss.*

While in the inner solar system active JFCs undergo substantial mass-loss/unit area due to a range of processes that include sublimation of $H_2O$, release of super-volatiles (CO and $CO_2$), release of organic and silicacious dust, outbursts, and nucleus splitting events. While some of these processes are, in essence, continuous, others are episodic. Some depend on the perihelion distance, $q$, and eccentricity, $e$ of the comet's orbit; others apparently do not. Some depend on the rotation state, which itself changes with time. Mass-loss in some comets may also be time-dependent in the sense that the mass-loss process itself slowly chokes off further mass-loss and the comet nucleus eventually becomes dormant or "dead" (Jewitt, 2004). If this process occurs on timescales that are much shorter than $\tau$ then, as outlined in the previous section, the JFC population may contain a significant fraction of inactive nuclei.

Since our knowledge and understanding of these processes is, at least in our opinion, rudimentary, we average their effects over the entire distribution and represent their combined mass-loss rate, averaged over many orbital periods and the entire population, by $\dot{m}$, the mass-loss rate *per unit surface area*. For the presumably dominant mass-loss process, $H_2O$ sublimation, this quantity is independent of the size of the object and can be represented by a continuous function. As a benchmark, assuming unit active fraction, Jewitt (2004) estimated $\dot{m}$ for a typical active JFC suffering mass-loss due to $H_2O$ sublimation at $-10^{-5}$ kg/m$^2$/s averaged over a single orbit. Mass-loss rates for other processes are extremely uncertain relative to the case of sublimation but may be substantial and they, unlike sublimation, may also be dependent on $r$. Belton *et al.* (2013) estimated in the case of 9P that the time-integrated mass-loss due to outbursts might approach half that of sublimation. Boehnhardt (2004) estimated that over its mean lifetime a short-period comet may lose 500 – 1000 m of equivalent radius to splitting events over its lifetime. For a not un-typical active fraction of 0.01, this is a time-averaged a mass-loss rate possibly as large as $-5 \times 10^{-8}$ kg/m$^2$/s, which is approximately half of the sublimational mass-loss rate that is quoted above. Nevertheless, we shall assume, since we have no other information, that the mass-loss rates/unit area due to all of these processes is independent of $r$ even though in the case of the latter two processes this is clearly an uncertain assumption.

How $\dot{m}$, averaged over $J(r)$ can be estimated from first principles is discussed in Appendix A. The timescale over which $\dot{m}$ is averaged must be much longer than the typical orbit period (~10 y, independent of size), the characteristic times between outbursts (~ 0.5 y, Belton *et al.,* 2013) and splittings (~ $10^2$ y, Chen and Jewitt, 1994; see also the review by Boehnhardt, 2004), and the spin evolution time ($10^2$-$10^3$ y) for a median sized nucleus [Jewitt, 2004]). All of these times are much shorter that the dynamical lifetime of $3.25 \times 10^5$ y (Levison and Duncan, 1997).

We presume that we can follow the rate of decrease in the effective radius, $\dot{r}$ of a "typical" active JFC nucleus, averaged over a timescale of $\sim 10^4$ y, with the following simple expression:

$$\dot{r} = (f/\rho)\dot{m} \tag{7}$$

where $\rho$ is the bulk density of the nucleus material and $f$ is an "active fraction" generalized to include mass-loss due to outbursts and splittings as well as sublimation. We view $f$ as an efficiency factor rather than an actual area. Note that if sublimation losses are dominant, then $\dot{r}$, like $\dot{m}$, should be largely independent of $r$ and depend only on the particular $a$ and $e$ of a comet's orbit.

The orbital parameters $a$ and $e$ of JFCs are not stable, suffering both gravitational and non-gravitational perturbations (Yeomans *et al.* 2004), however Levison and Duncan (1994) found that $T$ of the JFCs varied little over a $10^7$y numerical integration. They also found that $q$, which we expect to be the most significant orbital determinant of mass-loss, changed from orbits with q < 2.5 AU to orbits with q > 2.5 AU a median of some 10 times for a typical JFC during its dynamical lifetime. It is therefore seems clear that, even if we had all of the information available to calculate the evolution of the SFD by numerically following all of the known JFCs, it would be, in our opinion, a daunting and uncertain task. Some of the problems with such an approach are evidenced in recent work by Di Sisto *et al.* (2009) who followed the evolution of the JFC population directly with a numerical dynamical-physical model. It is because of the many problems latent in this technique that we have chosen to employ a statistical approach under the hypothesis that the JFC size frequency distribution, averaged over $\sim 10^4$ y, is in a steady-state with all the implications that has for detailed balancing of changes in orbital properties of the JFCs.

2.2 *Estimating the scattered disk size-frequency distribution from the JFCs.*

We have already made the basic assumption that mass-loss has no significant effect on the SFD in the Centaur population (while, nevertheless, recognizing that this assumption may have to be modified as future observations become available). It follows that $S(r)$ is related to the SFD the source population in the scattered disk, $SD(r)$, by an equation similar to (2) but without any distinction between active or inactive objects. We therefore write:

$$SD(r) = A\left[(1 + B/r)J(r) + C\frac{dJ(r)}{dr}\right] \tag{8}$$

where the coefficients are $A = 1/\alpha_{SD}\tau_S\alpha\tau$ ; $B = \tau\beta_0$; and $C = \tau\dot{r}$. Here $\alpha_{SD}$ is the injection rate from the scattered disk into the Centaur population and $\tau_S$ is the dynamical lifetime of the Centaur source population. The value of $A$ relates the number of JFCs to the number of source SDOs, $B$ controls the relative numbers of active and defunct nuclei, and $C = \tau \dot{r}$ is the characteristic loss in radius suffered by JFCs averaged over the whole population during their dynamical lifetime. For Jewitt's (2004) "canonical" JFC with $\dot{m}$ = -10$^{-5}$ kg m$^{-2}$ s$^{-1}$, $\rho$ = 500 kg m$^{-3}$, and an active fraction 0.05, we find $\dot{r}$ = - 2 10$^{-10}$ m s$^{-1}$. This would suggest a value for $C$ of - 2 km, which is probably a lower limit since most comets spend more time on orbits with larger $q$ than the orbit considered by Jewitt. It also implies that the source objects in the scattered disk may fall in a range that is a little larger than the $0.1 < r < 10\ km$ that we normally associate with JFCs, *e.g.,* $1 < r < 12$ km.

To estimate $SD(r)$ from Eq. 8 we calculate $J(r)$ and $\frac{dJ(r)}{dr}$ from the *observed* differential SFD, $J_{obs}(r)$, by correcting for observational selection effects. We then adjust parameters $A, B,$ and $C$ to fit what is already known observationally about the scattered disk SFD (see section 5 below).

### 3. $J_{obs}(r)$, the observed JFC size-frequency distribution.

As mentioned earlier, $J_{obs}(r)$ in its cumulative form has been the subject of a number of recent investigations (Meech *et al.* 2004, Tancredi *et al.,* 2006, Snodgrass *et al.,* 2011; Weiler *et al.,* 2011; Fernández *et al.,* 2013; [earlier work is well referenced in these papers]) and reviews *e.g.,* Lamy *et al.* (2004). The list of comets continues to slowly grow with time and reasons for disagreements on the sizes of particular JFCs in the different lists exposed. We have reviewed all of this material and generated our own list of 161 preferred effective radii (Table 1). Unlike the practice in much of the source material we quote the radii and their estimated uncertainties to the nearest 0.1 km, which we consider to be a realistic reflection of the overall level of accuracy that has been achieved. The information contained in Table 7 of Fernández *et al.* (2013), which is based on Spitzer IR observations, has, with the caveat noted above, been included as published. There is an overlap between the IR and Visual range observations in 39 of our comets and since two different observational techniques are involved we have investigated whether there are any systematic differences in the estimates. We find that while there are clear differences in the estimates for specific comets, a linear regression between the two types of estimate shows a linear trend with a slope of 0.94 and a variance of 0.74 km. We conclude that any systematic differences are minimal.

The effective radii of our 161 comet nuclei range from 0.2 to 15.4 km with the average at 1.9 km. While our approach to assigning radii is partially subjective, we note that there is much in common in the various lists and where major disagreements exist we have generally opted in favor of the smaller radius in the available estimates because of the chance of undetected coma activity. In cases where several of the estimates are in reasonable agreement we have taken a simple average. We also include in the list the best estimates of mean radius from spacecraft investigations. In the case of the Snodgrass *et al.* (2011) study, the individual radii that they used to construct their best estimate of the cumulative size distribution are not included in their paper as they used a Monte-Carlo technique to produce a range of possible radii matching the original observations for each comet. However, we have been very fortunate to receive from Dr. Snodgrass (private communication) a file of their "Monte-Carlo run 39," which is the basis of their final result. From this material we have been able to estimate individual radii and, just as significantly, an estimate of their observational uncertainties. This information has been folded into the data in Table 1. To demonstrate that the comets in our list fall within the definition of JFC membership (Eq. 1), we also include individual values of $T$ and other orbital information.

To show that our list of radii is reasonably compatible with the results of earlier studies we compare in Figure 1 a cumulative size distribution based on our list of 161 objects with the with the cumulative SFD Meech *et al.* (2004) (49 objects) and that of Snodgrass *et al.* (2011) (86 objects). They are all clearly similar in shape after consideration of the number of objects involved. The slope of our cumulative SFD between $1 < r < 10$ km is -2.19, which should be compared with the values derived by Fernández *et al.* (201), Weiler *et al.* (2011), Snodgrass *et al.* (2011), Tancredi *et al.* (2006), and Meech *et al.* (2004) quoted above. We believe that it is significant to note that neither the cumulative SFD based on the data from Table 1 nor that displayed in Figure 1 of Snodgrass *et al.* (2011), or their Figure 9, which in reality is a probability map of many cumulative size-distributions from many Monte-Carlo runs, are linear when displayed on a log-log plot indicating that they do not truly represent power-law distributions. This why the reported slopes are so sensitive to the range of $r$ over which they are estimated. These slopes are, in our opinion, mainly useful for anecdotal purposes and we shall continue to use them for comparative purposes. Nevertheless, they probably have little physical significance. We refer the reader to the paper by Stumpf and Porter (2012) on the value of power-law fits in scientific applications and the conditions under which they might usefully be invoked. In this work we will avoid cumulative distributions as much as possible (but not entirely) and focus as much as possible on differential SFDs. Since we will be using power-law slope estimates for making comparisons it should also be noted that while for a true power-law distribution the difference between the cumulative and differential slopes is unity, for real distributions that are not exactly power-laws the difference is only approximately unity.

3.1 *Orbital properties of the observed JFCs.*

In Figure 2 we show plots of $r$ versus $a$ and $e$ to illustrate the degree to which $J_{obs}(r)$ depends on these orbital parameters. In Figure 3 we plot of $r$ versus $T$ to demonstrate that the comets in Table 1 fall within the limits set above for JFC membership and to show that there is no indication that $J(r)$ depends on $T$. Since the number of comets is relatively small we also show in Figure 2 plots of $J_{obs}(r)$ for two ranges each of $a$ and $e$, one including all objects below the median value and the other above. It is apparent that the peaks of the observed distributions (Figures 2c and 2d) move to higher values of $r$ as $a$ increases and $e$ decreases, *i.e.,* there is a dependence of $J_{obs}(r)$ on $a$ and $e$. This could, in both cases, simply be the result of observational selection since it is obviously harder to find small objects on orbits that place the object further from the sun. This is discussed further in Appendix A.

3.2. *The observed differential size-frequency distribution.*

It is usual to display the differential SFD in its incremental form as a binned histogram (an insightful look at the relationship of differential and incremental forms of such distributions is presented by Colwell, 1993) and in Figure 4 we compare our $J_{obs}(r)$ incremental distribution to an earlier one by Meech *et al.* (2004). The interpretation of the shape of binned incremental distributions is sometimes subject to discussion because they can often display a substantial dependence on what bin size in chosen. We choose to avoid this issue by employing a *probability index* (PI) technique that we have used earlier in describing the initiation time of CN shells from the nucleus of 1P/Halley as it spins (Belton *et al.,* 1991). With this technique we build up a smoothed approximation to the differential distribution from normalized Gaussian functions that represent the probability that the effective radius of an observed object is at $r$, based on its most probable observed value and its associated uncertainty. The probability index for the distribution is:

$$PI(r) = \sum_i^n \frac{1}{\sqrt{2\pi}\sigma_i} e^{-0.5\left(\frac{r-r_{0i}}{\sigma_i}\right)^2} \qquad 9$$

where $r_{0i}$ is the probable value of the effective radius of nucleus $i$, $\sigma_i$ is its uncertainty, and $n$, is the number of nuclei. $PI(r)dr$, can be thought of as the number of "virtual" objects between $r$ and $r + dr$. With this approach uncertainties associated with the choice of bin size are removed and all of the information that we have about effective radii and their uncertainty is used. This approach has the benefit that the resulting distribution is continuous and can be sampled at any value of $r$. Also, it can be seen intuitively that as the observed number of objects is increased and as the observed radii become more precisely defined, the derived $PI$ will approach the actual differential distribution. In Figure 5 we compare $J_{obs}(r)$ calculated in this way and plotted at 0.2 km intervals, with the binned histogram version shown in Figure 4. In Table 2 we list, sampled at 0.2 km intervals, values of $PI(r)$ ($\equiv J_{obs}(r)$) and an estimate of its formal uncertainty. Further discussion of uncertainties is reserved for section 5.

## 4. The size-frequency distribution of the active JFC source population.

To obtain $J(r)$ from $J_{obs}(r)$ we need to quantify the effects of observational selection on $J(r)$ with a model of the process. Such a model was proposed by Meech *et al.* (2004) and, although it led to the interesting result that the observed SFD of the JFCs was incompatible with a power-law in general, their observational selection model appears to have received little more than passing reference in subsequent critical discussions. In fact, in two subsequent studies of the SFD of JFCs quite different approaches to selection effects on this population are offered: In Tancredi *et al.* (2006) it is argued that "…we concentrated our analysis in a sub-sample that corresponds to a more homogeneous population in the dynamical sense (the JFCs) with a high degree of completeness (comets with $q < 2 - 2.5\ AU$) therefore we do not have to debias the sample…", and in Snodgrass *et al.* (2011) they state after presenting relevant arguments: "In this study we therefore assume that the observed cometary nuclei approximate a randomized sample of the JFC population." Here we simply note that, while these arguments may have some validity, the Meech *et al.* (2004) model, as applied numerically to various assumed initial size distributions, has a very strong effect particularly towards smaller sizes. This is demonstrated graphically in their Table 12, which shows the effects of applying their observational selection model to 20 different initial SFDs. At 1 km effective radius corrections to the observed number of objects can amount to as much as a factor of 10.

    According to Meech *et al.* (2004) the "best" of their models has the following attributes: *(i)* The assumption of the presence of activity within 3 AU, *(ii)* a $\Delta$ effect that makes some comets harder to detect when closer to the Earth, and *(iii)* realistic detection limits. We have studied their application of the model to nine of the cases (D, L, Lb, Lc, N, Ob, Oc, P and Pb) in their Table 12. The first seven are power-law initial distributions with a wide range of slopes at values -1.5, -2, -2.5, -3, -3.5, -4, and -4.5; the latter two are "ramped" and truncated power-laws with their own distinctive shape. This was done by digitizing the relevant figures in their Table 12. The remarkable result of this exercise, at least to us, was our finding that the correction function is effectively the same in all cases, *i.e.,* with the Meech *et al.* (2004) model, the shape of the observational correction function (OCF) is independent of the shape of the initial distribution. This finding is not mentioned in the Meech *et al.* (2004) paper and we consider it a significant result since it allows us to apply a correction factor directly to $J_{obs}(r)$ in order to estimate $J(r)$. The OCF we have deduced is shown in Figure 6 and listed in Table 2 at 0.2 km intervals. Between 0.2 < r < 5 km the OCF is the average of the first seven of the cases noted above since they span the full range of effective radius. The function shows an slightly undulating structure with effective radius but because the result is strictly numerical it is not clear from the discussion in Meech *et al.* (2004) what the physical origin of these undulations is. In what follows we simply accept the results of Meech *et al.* (2004) on the OCF rather than attempt a separate calculation

of our own. Cases P and Pb, which only span half the range, were used as a check on our conclusion that the OCF is independent of the shape of the initial SDF used. Beyond r = 5 km we have extended the function linearly until it reaches unity near r = 7. Beyond this we have made the admittedly uncertain assumption that all the larger JFCs have been found (for example, the large nucleus (effective radius 7 km) of 162P/Siding-spring was found as recently as 2004) and put the correction factor identically equal to one.

In Figure 7 we show our estimate for $J(r)$ with uncertainties based on the number of virtual nuclei in 0.2 km bins. To estimate the *total* number of active JFCs in the range 0.2 < r < 15 km we must not only account for the active nuclei with size estimates (~ 833 based on our estimate of $J(r)$) but also those discovered active comets that meet the criteria in Eq. 1 and that are currently without a size estimate (there are 276 such objects that have been observed). Here we have found it necessary to make the assumption that the size-frequency distribution of the JFC nuclei without size estimates is the same as that of the nuclei with size estimates. With this assumption we can use the same OCF for both populations. While this is obviously a questionable assumption and, at best, can only be approximately correct we anticipate that it will provide a rough estimate to guide future work. With this assumption we find that there should be approximately 2300 objects in the active JFC population that meet our criteria.

To estimate the total population of active plus defunct JFC nuclei this number will need to be increased to account for the "dead" or dormant nuclei (see section 6 below). Evidently, only a small fraction of active JFCs (~ 7%) have had their radii estimated so far. This is in accord with the finding by Fernández *et al.* (2013), based on discovery statistics, that the known JFC population is far from complete. Other estimates of the total number of JFCs have been made. Most recently, Di Sisto *et al.* (2009) using dynamical-physical modeling of the evolution of the JFC population find 450 ± 50 JFCs with r >1 km while our SFD distribution suggests ~ 1173 nuclei in this range of $r$. For comparison with other published work we show in Figure 8 the cumulative size distribution implied by $J(r)$. Between 1 < r < 10 km the cumulative power-law slope of the distribution is -2.82, which is considerably steeper (by about one unit) than that measured for the $J_{obs}(r)$ distribution or that predicted for strength-less nuclei (-2.04) by O'Brian and Greenberg (2003). Also, between 3 and 6 km effective radius, there is no obvious sign of a possible feature, noted by Fernández *et al.* (2013), in the curve (other than noise).

## 5. The scattered disk TNO size–frequency distribution

*5.1 Handling the mass-loss term in Eq. 8*

Given $J(r)$ and $\frac{dJ(r)}{dr}$, it should be a straight forward matter to use Eq. 8 to estimate $SD(r)$ in the range 0.2 < r < 15 km by choosing $A, B$ and $C$ so that $SD(r)$ matches existing observations of the TNO size distribution, *i.e.*, the Schlichting *et al.* (2012, 2013) occultation constraint and the Fuentes *et al.* (2010) "hot" SFD distribution (used as a surrogate for the scattered disk population) that ranges between effective radii of ~15 to 400 km. But, because $J(r)$ is determined from a relatively small sample of objects, our estimate for $J(r)$ unavoidably contains noise that appears as undulations or "wiggles" on the distribution (Figure 7). By simply numerically differentiating $J(r)$ to obtain $\frac{dJ(r)}{dr}$ these "wiggles" will be amplified in any estimate of $SD(r)$. To filter out this noise we have used a numerical procedure based on the following identity to calculate $\frac{dJ(r)}{dr}$:

$$\frac{d(Log10(J(r)))}{dr} = \frac{Log10(e)}{J(r)} \frac{dJ(r)}{dr} \qquad (10)$$

Figure 9 shows $Log_{10}(J(r))$ plotted against both r (log-linear) and $log_{10}(r)$ (log-log) and, except for the above mentioned wiggles, it is clear that $Log10(J(r))$ can be represented by a linear regression over the range 0.6 < r < 10 km, *i.e.*,

$$Log10(J(r)) \approx -0.4358\, r + 3.0023 \qquad (11)$$

It is also clear from the curvature in the log-log plot of the right panel in Figure 9 that $J(r)$ is not a power-law size distribution in $r$ (however, in section 6 we shall find it expedient to use an effective power-law fit to various distributions for comparative purposes). Adopting Eqs. 10 and 11 we have:

$$\frac{dJ(r)}{dr} \approx -J(r) \qquad (12)$$

and Eq, 8 can be rewritten

$$SD(r) \approx A'(1 + B'/r)J(r) \qquad (13)$$

where $A' = A(1 - C)$ and $B' = \frac{B}{1-C}$. Eq. 13 should hold for at least for 0.6 < r < 10 km and, possibly, a much wider range (Note that $C$ is always negative since it refers to mass-loss). It is interesting that, if defunct nuclei are ignored, Eq. 13 justifies the assumption, often made in earlier work, that the shape JFC size distribution is similar to that of source objects in the TNO population. However, if a defunct sub-class exists within the JFC population this justification is lost.

*5.2 Fitting $SD(r)$ to TNO observations.*

Figures 10 and 11 are cumulative plots (number of objects/square degree with effective radius $> r$) for the TNO population near the ecliptic partially digitized from Figure 6 in Schlichting *et al.* (2013). It shows the Schlichting *et al.* (2012) occultation point and its 95% upper and lower confidence limits together with the observations of hot TNOs (shown schematically) for objects larger than ~ 15 km radius from Fuentes *et al.* (2010). To convert the magnitudes of Fuentes *et al.* (2010) to radii we have assumed an albedo of 0.04 and a standard distance of 42 AU. Also shown is the 95% confidence upper limit on the number of small KBOs from the TAOS survey reported by Zhang *et al.* (2013). Superposed on Figure 10 are our results for a cumulative version of $SD(r)$ computed with $A' = 300; B' = 0$, *i.e.*, the case of no defunct nuclei. Note that although there is considerable uncertainty near $r = 10\ km$ in our data, the uncertainty falls off rapidly to smaller radii and so the choice of $A'$ is limited. We use the following artifice to choose $A'$: we fit a power-law to $SD(r)$ in the range 1 < r < 6 km (shown in the figure) and use its extension to larger radii to match the hot TNO population that has a very similar slope. We place an uncertainty of ± 100 on the value of $A'$. With this fit to observed TNOs, our data does not pass through the 95% confidence limits of the Schlichting *et al.* (2012) occultation point but it does fall below the TAOS limits and fit smoothly to the Fuentes *et al.* (2010) data. Either the assumption of no defunct nuclei in the JFC population is incorrect or the Schlichting point refers to some other population than the source objects for the JFC nuclei and there are ~ 50 times fewer objects in TNO population than their estimate suggests.

In Figure 11 we show the case in which defunct nuclei are included as a sub-class of the JFCs. This, our best fit, is for $A' = 500, B' = 6$. It allows a smooth transition between $SD(r)$ and the observed TNO hot population, matches the 1σ limit of the Schlichting occultation point, and avoids (barely) violating the 95% confidence upper limit set by the TAOS survey. It also implies that a large number of defunct nuclei exist in the JFC population (see below). In Figure 12 we show the differential SFD for the total JFC population that corresponds to this fit and that shows the relative influence of active and defunct nuclei in the JFC population on the shape. Our estimate for the differential power-law index for the overall TNO population in the range 1 < r < 10 km is -4.5 +/- 0.5 (cumulative slope is -3.24). Here the uncertainty estimate for the differential slope is not a formal error but a conservative visual estimate that takes into account the curved nature of the distribution on a log-log plot.

6. Discussion

A significant result of this study is to demonstrate that the cumulative slope of the observed SFD of active Jupiter family comets, which has been a focus of several earlier studies (Fernández *et al*, 2013; Weiler *et al.,* 2011; Snodgrass *et al.,* 2011; Tancredi *et* *al.,* 2006; and Meech *et al.*, 2004), is far from equivalent to the cumulative slope of the hot TNO population (-2.91). We find that in order to estimate the latter slope from the former it is essential account for observational selection in the discovery of active JFCs and to include the effects of devolatization of comet nuclei into a defunct state. In order to match the observational constraints on the small end of the hot TNO distribution we find that each of these processes adds approximately -0.6 units to the cumulative power-law index of the observed size distribution of JFCs. Mass-loss, as indicated earlier by Weissman and Lowry (2003) (see also Lowry *et al.,* 2008), is found to be a much less significant factor. Note that while the model for the OCF used in section 4 and the parameterization of devolatization by $\beta_0/r$ in Eq. 5 produce a size distribution that can fit TNO observations, they nevertheless carry substantial uncertainties and are largely untested.

With regard to the question about the nature of cometary nuclei posed at the beginning of this paper this study has produced three significant results. First, there must be a large number of defunct cometary nuclei in the JFC population and, except for a few in the NEO population (Fernández *et al.*, 2002) and a few among the ACO population (Licandro *et al.,* 2008), most have yet to be conclusively identified. This result implies a relatively short devolatization timescale for active comets. The parameter $B' = \tau\beta_0/(1 - \tau\dot{r})$ contains information on this - albeit combined with the effects of mass-loss. With $B' = 6$, $\tau = 3.5x10^5$ $y$ and assuming that, averaged over the entire population, a typical nucleus loses 250 m of its radius over $\tau_{dv}$ the devolatization time, we find that $\tau_{dv} = 4.7x10^4$ y for a 1 km radius nucleus. This is about a factor of four less than a previous estimate by Jewitt (2004), which is not unreasonable since the latter estimate only considered radius loss to $H_2O$ sublimation. That the number of defunct nuclei must be large and concentrated at smaller sizes is indicated in Figure 12. In section 4 we found that between 0.2 and 15 km radius there should be approximately ~2300 active nuclei in JFC population, some 14 times more than have had their radii measured, and with $B' = 6$ and $C = -250$ m we find that there should be some 9 times more defunct nuclei, *i.e.,* roughly 22,000 with the vast majority having radii < 1km.

J. Scotti (2013, private communication), has suggested that possible check on this might come from the recent expansion of the MPC orbit catalog now at more than 622,367 entries (October, 2013) with over 28,405 that fall with the JFC region of the $(a, e)$ plane (Eq.1). Some of these objects are as faint as H ~30 mag (effective radius = 3.6 m). Although the Jupiter Trojans, Hildas, Cebele, some NEO and main belt asteroids and presumably some fraction of undiscovered active comets, fall in this region, we surmise that the majority of those with $e$ > 0.55, are, in fact, defunct comet nuclei. In this latter interval there are 167 observed active JFCs and 931 ACOs giving a ratio of ~ 6, not far from the ratio implied by our fit to the TNO objects. If we assume that defunct comet nuclei are distributed the same way throughout the full range of eccentricity, *i.e.,* 0 < e <1, and this ratio applies to the populations after correction for observational selection, we estimate that the defunct comet population could be near ~2400 in the current MPC catalog and that the *total* population of defunct JFC comet nuclei could be as large as ~12,600. This is ~0.6 times the estimate from our fit to the TNO population described above. Since our assumption that all the objects for e > 0.55 in the MPC catalog may not be correct it appears that the number of defunct nuclei predicted by our fit to the TNO data is probably excessive. Possibly the size-dependence that we assumed for the devolatization process (Eq.5) is incorrect or another loss process, *e.g.*, disintegration of nuclei, may need to be accounted for. The average and median effective radii of objects in the MPC catalog with JFC orbits and e > 0.55 (assuming an albedo of 0.04) is 83 and 40 m respectively which shows that the MPC entries cover the range of sizes considered in this paper and are therefore relevant. We have looked at the SFD for these objects and find a cumulative power-law index of -1.5 for objects between 1 and 10 km effective radius, considerably shallower than the cumulative index found in section 3 for the observed active JFCs (-2.19). This slope, while of interest for comparative purposes, needs to be corrected for the (currently) unknown effects of observational selection. However, we note that it is numerically the same slope that was estimated for defunct cometary nuclei in the JFC population by Whitman *et al.* (2006) by extrapolation from defunct nuclei in the NEO population that does include a correction for observational selection. In contrast, and according to Eq.4 and the results in section 4, we would expect the intrinsic cumulative power-law index of defunct comets to be around -3.8. There are clearly several conflicts arising between our predictions for the defunct comet population and the little that is currently known about them that deserves further consideration in future studies.

   A second finding is the steepness of the slope of the differential SFD between 1 and 10 km. With a differential power-law index of -4.5 +/- 0.5 (cumulative -3.24) characterizing this size range we are far from the expectations of a fully relaxed collisional evolution scenario that should give a differential slope of – 3.5. The cumulative slope for hot TNOs (-3.24) inferred from the JFCs between 1 and 20 km is close to the cumulative slope seen in the TNO hot population from 20 - 400 km radius (-2.91) and there is little evidence for a significant "break" in the combined distribution in the near 10 km radius due to collisional evolution as discussed by Schlichting *et al.* (2013) although the uncertainties in the JFC curve could allow a small inflection near 10km effective radius. The distribution deduced from the JFCs does not have the same characteristic shape that is seen in Schlichting *et al's* (2013) evolutionary calculations – a low slope (index ~ 2) starting in at ~10 km and then steepening to a slope of -5 to 6 at smaller radii - where this steep slope is apparently a reflection of the assumed initial SFD at the end of runaway growth. Evidently our results represent a new observational constraint on the SFD that results from the runaway growth phase. Also, if the slope of the TNO hot population is the signature of primordial objects unaffected by collisional evolution as stated by Schlichting *et al.* (2013), then we must conclude that the inferred objects down to ~1 km radius, *i.e.,* cometary nuclei, are to a large extent similarly unaffected. This our strongest evidence that the nuclei that have been observed by spacecraft are in fact primordial accreted objects largely unaffected by collisional evolution.

The third finding is that the cumulative distribution does turn over to a much lower slope at r < 1 km. This could be a break in the slope that signifies the onset of the effects of collisional evolution and could, possibly, provide a constraint on fragmentation models.

Finally, it should be stressed that the number of comet nuclei for which we have reliable radii is still small at 161. It is clear from this study and other considerations that our sample is still far from complete. More high quality observations and discoveries are obviously needed.

## 7. Acknowledgements

This work was partially supported by the Stardust-NExT and EPOXI missions under contract NNM07AA99C at the University of Maryland and at Cornell University under agreement 51326-8361 and we thank the Principal Investigators Drs. J. Veverka and M.F. A'Hearn for their continued support and encouragement. We thank Drs. K. Volk and N. Samarasinha for valuable comments on the first draft of the paper. We also thank Dr. Y. Fernández for permission to use his data on comet radii before final publication and Dr. C. Snodgrass for comments on an early draft of the paper and sharing unpublished results that greatly improved this work. We also thank Jim Scotti for providing some orbital elements and his suggestions regarding the MPC asteroid orbit catalog. This work also benefitted from discussions within the PSI comet and Tucson asteroid lunch groups and with Dr. K. Volk. Two anonymous referees also provided comments that led to substantial improvements in the paper.

## Appendix A. Estimating mass-loss in the JFC population

In section 2 we estimated what is probably a lower limit to parameter $C$ (Eq. 8), which is the product of the dynamical lifetime of the JFCs with the average rate at which $r$ decreases. It is therefore proportional to the mass-loss rate/unit area. In principle this parameter can also be estimated as an average over the whole population of JFCs. To do this we need to do a more detailed derivation of Eq. 8 noting that the mass-loss suffered by a nucleus depends on $a$ and $e$. In this derivation we ignore the complication of defunct nuclei, *i.e.,* we put $\beta_0 = 0$.

Let $J(a,e,r)$ and $S(a,e,r)$ be the differential SFD functions for objects of effective radius r such that $J(a,e,r)$ is the SFD for those objects in the JFC population characterized by orbits with elements between $a$ and $a + da$ and $e$ and $e + de$. Similarly $S(a,e,r)$ is the SFD of the source population that injects objects on to orbits with elements between $a$ and $a + da$, and $e$ and $e + de$. We have therefore:

$$S(r) = \iint S(a,e,r) \, dade \quad \text{and} \quad J(r) = \iint J(a,e,r) \, dade \qquad \text{A1}$$

So, following the argument in section 2.2, Eq. 8 can be written:

$$SD(r) = A(J(r) + \tau \iint \dot{r}(a,e) \frac{dJ(a,e,r)}{dr} dade) \qquad \text{A2}$$

since $\dot{r}$ is independent of $r$.

In section 3.1, we point out that while the distributions displayed in figures 2c and 2d show that $J_{obs}(r)$ has a dependence on both $a$ and $e$. The shifts seen in the figures between higher and lower values of $a$ and $e$ are in the sense that would be expected to result from the effects of observational selection. Here, with a limited supply of observations, we have little choice but to assume that this is the actual reason for the shifts and that when observational selection effects are accounted for, $J(r)$ will become independent of $a$ and $e$. With this assumption A2 becomes:

$$SD(r) = A(J(r) + \tau[\iint \dot{r}(a,e) \, dade] \frac{dJ(r)}{dr}). \qquad \text{A3}$$

or, $C \sim \tau \iint \dot{r}(a,e) \, dade$ \qquad A4

This allows, given a physical model for mass-loss, computation of an estimate of $C$ over the observed population of JFCs.

22# References

Basilevsky, A.T., and Keller, H.U., 2007. Craters, Smooth Terrains, Flows, and Layering on the comet nuclei. Solar Sys. Res. 41, 109-117

Belton, M.J.S., et al., 2013. The origin of pits on 9P/Tempel 1and the geologic signature of outbursts in Stardust-NExT images. Icarus 222, 477-486.

Belton, M.J.S., 2010. Cometary activity, active areas, and a mechanism for collimated outflows on 1P, 9P, 19P and 81P. Icarus 210, 881-897.

Belton, M.J.S., Thomas, P., Veverka, J., Schultz, P., A'Hearn, M.F., Feaga, L., Farnham, T., Groussin, O., Li, J-Y., Lisse, C., McFadden, L., Sunshine, J., Meech, K.J., Delamere, W.A., and Kissel, J., 2007. The Internal Structure of Jupiter Family Cometary Nuclei from Deep Impact Observations: The "talps" or "Layered Pile" Model. Icarus 187, 332-344.

Belton, M.J.S., W. H. Julian, A. J. Anderson, and B. E. A. Mueller., 1991. "The Spin State and homogeneity of Comet Halley's Nucleus," Icarus 93, 183-193.

Boehnhardt, H., 2004. Split comets. In "Comets II." Festou, M.C., Keller, H.U., and Weaver, H.A. (Eds.). Univ. of Arizona Press, Tucson, 301-316.

Brasser, R., and Morbidelli, A., 2013. Oort cloud and scattered disk formation during a late dynamical instability in the solar system. Icarus 225, 40-49.

Brownlee, D. *et al.* 2004. Surface of Young Jupiter Family Comet 81 P/Wild 2: View from the Stardust Spacecraft. *Science* 304, 1764-1769

Charnoz, S., and Morbidelli, A. 2003. Coupling of dynamical and collisional evolution of small bodies. An application to the early ejection of planetesimals from the Jupiter-Saturn region. Icarus 166, 141-156.

Charnoz, S., and Morbidelli, A. 2007. Coupling of dynamical and collisional evolution of small bodies II. Forming the Kuiper Belt, the scattered disk and the Oort cloud. Icarus 188, 468-480.

Chen, J. and Jewitt, D.C. 1994. On the Rate at Which Comets Split. Icarus, 108, 265-271

Colwell, J.E., 1993. Power-law confusion: You say incremental, I say differential. Lunar Plan. Sci Conf. XXIV, 325-326.

**Figure captions**

Figure 1. Cumulative size-frequency distributions (CSFDs) for JFCs. The solid line is the CSFD for the data in Table 1. The open circles are a digitized version of the CSFD in Figure 9 of Snodgrass *et al.* (2011). The filled squares are a digitized version of the CSD in Figure 6 of Meech *et al.* (2004). The slope for 1 < r < 10 km in the Table 1 data is -2.19.

Figure 2. Distributions of $r$ against $a$ and $e$. (*Top left*) $r$ against $a$. The vertical dashed line divides the distribution at the median value between low and high values of $a$. (*Top right*) $r$ against $e$. The vertical dashed line divides the distribution at the median value between low and high values of $e$. (*Bottom left*) $r$ distributions for low (solid) and high (dashed) values of $a$. The distribution is marginally shifted towards higher values of $r$. (*Bottom right*) $r$ distributions for low (solid) and high (dashed) values of $e$. The distribution is marginally shifted towards smaller values of $r$. The shifts in the position of the maximum in the bottom graphs are in the sense expected to result from observational selection effects.

Figure 3. Distribution of $r$ against $T$. There is no apparent dependence of the $r$ distribution with $T$.

Figure 4. Binned differential SFDs for observed JFCs. The bin size is 0.5 km and the data are plotted in the same manner as they appear in Meech *et al.* (2004). The data in Table 1 is represented by a solid line and includes 161 nuclei; the data in Figure 6a of Meech *et al.* (2004) is represented by the dashed line and includes 49 nuclei. After accounting for the smaller number of nuclei available to Meech *et al.* (2004), the shapes of the SFDs are quite similar.

Figure 5. (*Solid line*) Probability Index ($J_{obs}(r)$) for the data in Table 1 sampled every 0.2 km as compared to a binned histogram of the same data. Both curves have been normalized so that the area under the curve is equal to the number of nuclei (161). Unlike Figure 4, which was plotted to match the way in which Meech *et al.* (2004) presented their data, this figure plots the binned histogram so that the center of the first bin (0.5km wide) is placed at $r = 0.25\ km$. The shapes of both distributions are similar.

Figure 6. Observational correction function (OCF) based on the model in Meech *et al.* (2004). (*Left panel*) Measurements and average curve from seven of different initial distributions (D, L, Lb, Lc, N, Ob, Oc) digitized from Table 12 of Meech *et al.* (2004). It is clear that, except possibly at the very largest effective radius, the OCF is independent of the initial population. (*Right panel*) Final correction curve with a subjective extension beyond 5 km radius (see text).

Figure 7. The probability index SFD, $J(r)$, for JFCs after correction for observational selection. The $J(r)$ function is plotted at 0.2 km centers and the uncertainty bars at selected radii are based on Poisson statistics of the number of virtual nuclei in 0.2 km bins in the original $J_{obs}(r)$ and then amplified by the observational selection correction factor. The low-frequency undulations on the curve are also "noise" due to the limited number of nuclei used in calculating the probability index. These undulations can also be seen in Figure 8.

Figure 8. Scaled cumulative size distribution of JFCs after correction for observational selection. The power-law trend line is fitted between 1 < r < 10 km and has a slope of -2.82.

Figure 9. (*Top panel*) Log-linear plot of $J(r)$. The curve, except for "noise" undulations, is effectively linear between 0.6 and 10 km with a slope of -0.4358. Below 0.6 km and beyond 10 km there are few observed nuclei (5 and 2 respectively) and the curve is poorly defined. The undulations in the curve are noise due to the limited number of nuclei in the sample used to calculate the probability index. (*Bottom panel*) Log-log plot of $J(r)$ over the full range of observed effective radii. The curvature seen over the range from 0.6 to 10 km indicates that $J(r)$ is not well represented by a power-law. The linear trend line fitted to the points between 1 and 10 km has a slope of – 3.58.

Figure 10. Cumulative plot of the number of objects in the TNO population near the ecliptic. The cumulative form of $SD(r)$ for A' = 350, B' = 0 is superposed (solid line) and is plotted from r = 0.2 to 20 km. The curve is very uncertain between 10 and 20 km since only two comets define this part of the curve. This fit represents the case for no defunct nuclei in the JFC population. The match to the schematic representation of the data of Fuentes *et al.* (2010) on the hot TNO population is done by extending the power-law trend line (dots), defined by points in the range r = 1 to 6 km of the cumulative SD(r) curve (see text for details). The power-law index of the trend line is -2.66. The line marked with + is an upper limit on the total number of TNOs from the TAOS survey (Zhang *et al.* 2013). The point at r = 0.25 km is the "occultation point" of Schlichting *et al.* (2013) with 95% confidence lines (open diamonds). It is not possible to simultaneously fit the Schlichting occultation point and match the slope of the hot TNO population if $B' = 0$. *i.e.,* if defunct nuclei are excluded. Note also the transition to a lower slope near 1 km radius. See section 5.2 for a discussion.

Figure 11. Cumulative plot of the number of objects in the TNO population near the ecliptic. The cumulative form of $SD(r)$ for A' = 500, B' = 6 is superposed (solid line)

and is plotted from r = 0.2 to 20 km. However the curve is very uncertain between 10 and 20km since only two comets define this part of the curve. This is the case that includes defunct nuclei in the JFC population. The match to the schematic representation of the data of Fuentes *et al.* (2010) on the hot TNO population is done by extending the power-law trend line (dots), defined by points in the range r = 1 to 6 km of the cumulative SD(r) curve (see text for details). The power-law index of the trend line is -3.24. The line marked with + is an upper limit from the TAOS survey (Zhang *et al.* 2013). The point at r = 0.25 km is the "occultation point" of Schlichting *et al.* (2013) with 95% confidence lines (open diamonds). This is our "best fit" to the TNO data since it simultaneously satisfies all three of the available observational constraints. The cumulative SD(r) slope roughly matches the slope of the Fuentes *et al.* (2010) data on the hot TNO population, it avoids the TAOS upper limit and passes through the 1 sigma uncertainty associated with the Schlichting occultation point. Note also the transition to a lower slope near 1 km radius. See sections 5.2 and 6 for a discussion.

Figure 12. The differential SFD for the JFC comets based on the A' = 500, B' = 6 fit to TNO data (Figure 11). The dashed line is for the total population while the solid line shows the active population after correction for observation selection effects (compare Figure 7). The difference between the two curves measures the (large) population of defunct comets. The uncertainty bars represent the uncertainties in J(r) propagated to SD(r) (see caption to Figure 7). They do not include possible systematic uncertainties in the OCF and $\beta_0$/r.

# TABLE 1

Table 1. Effective radii for 161 JFCs. The radii and uncertainties (here assumed symmetric) are given to the nearest 0.1 km, even though much of the original observational data is published to the nearest 0.01 km and uncertainties are often given as asymmetric. This list was formulated during a review of the literature (see text) and our assessment is that, except for some of the spacecraft data, ±0.1 km is a more appropriate level of uncertainty for our purposes. The values of the uncertainties are either based on the Fernández *et al.* (2013) original listing or on the Monte-Carlo run 39 (Snodgrass *et al.*, 2011), the results of which were kindly provided to us by Dr. C Snodgrass. For comets not included in the Snodgrass *et al.* (2011) run 39 or the Fernández *et al.* (2013) list, with the exception of 9P, 19P, 81P and 103P that were the subject of spacecraft investigations, we used an approximate linear correlation between effective radius and uncertainty based on comets that were included in run 39. We also display values of the orbital elements, primarily obtained from the JPL Horizons listings. The Tisserand parameter was calculated relative to Jupiter. For a few of the more recently discovered comets the elements where derived from CBAT notices or provided by an individual observer. The average effective radius in this listing is 1.9 km. Column 5 indicates the references used in arriving at an estimate to the effective radii and its uncertainty. The references are: a = Fernández *et al.* (2013); b = Snodgrass *et al.,* (2011) – MC run 39; c = Meech *et al.* (2004); d = Tancredi *et al.* (2006); e= Brownlee *et al.* (2004); f = Thomas *et al.* (2013a); g = Thomas *et al.* (2013b).

| Number | Name | Radius (km) | Uncertainty (km) | Ref. | a (AU) | e | I (deg) | q (AU) | P (y) | T |
|---|---|---|---|---|---|---|---|---|---|---|
| 1 | 2P | 2.3 | 0.6 | c,d | 2.21 | 0.848 | 11.78 | 0.34 | 3.29 | 3.03 |
| 2 | 4P | 1.8 | 0.3 | c,d | 3.85 | 0.567 | 9.03 | 1.67 | 7.54 | 2.75 |
| 3 | 6P | 2.2 | 0.2 | a | 3.50 | 0.613 | 19.52 | 1.35 | 6.53 | 2.71 |
| 4 | 7P | 2.6 | 0.2 | a | 3.43 | 0.635 | 22.31 | 1.25 | 6.36 | 2.68 |
| 5 | 9P | 2.8 | 0.1 | g | 3.12 | 0.517 | 10.53 | 1.51 | 5.52 | 2.97 |
| 6 | 10P | 4.2 | 0.7 | b,c,d | 3.06 | 0.536 | 12.03 | 1.42 | 5.37 | 2.96 |
| 7 | 11P | 0.6 | 0.1 | a | 3.44 | 0.539 | 13.46 | 1.58 | 6.37 | 2.85 |
| 8 | 14P | 3.0 | 0.2 | a | 4.24 | 0.358 | 27.94 | 2.72 | 8.74 | 2.72 |
| 9 | 15P | 0.9 | 0.1 | a | 3.48 | 0.721 | 6.82 | 0.97 | 6.50 | 2.62 |
| 10 | 16P | 0.7 | 0.1 | a | 3.35 | 0.563 | 4.26 | 1.47 | 6.14 | 2.87 |
| 11 | 17P | 1.6 | 0.2 | b,c,d | 3.62 | 0.432 | 19.09 | 2.06 | 6.89 | 2.86 |
| 12 | 19P | 2.5 | 0.2 | g | 3.61 | 0.625 | 30.33 | 1.35 | 6.85 | 2.56 |

| | | | | | | | | | |
|---|---|---|---|---|---|---|---|---|---|
| 13 | 22P | 2.2 | 0.2 | a | 3.46 | 0.545 | 4.73 | 1.57 | 6.42 | 2.87 |
| 14 | 24P | 0.9 | 0.1 | d | 4.08 | 0.705 | 11.75 | 1.21 | 8.25 | 2.51 |
| 15 | 26P | 1.9 | 0.2 | b,c,d | 3.02 | 0.640 | 22.42 | 1.09 | 5.24 | 2.81 |
| 16 | 29P | 15.4 | 3.1 | c | 6.00 | 0.043 | 9.37 | 5.74 | 14.71 | 2.98 |
| 17 | 30P | 1.0 | 0.1 | d | 3.78 | 0.501 | 8.12 | 1.88 | 7.34 | 2.84 |
| 18 | 31P | 1.7 | 0.1 | a | 4.25 | 0.194 | 4.55 | 3.42 | 8.75 | 2.99 |
| 19 | 32P | 2.4 | 0.2 | a | 4.26 | 0.570 | 12.93 | 1.83 | 8.80 | 2.67 |
| 20 | 33P | 1.2 | 0.1 | a | 4.03 | 0.462 | 22.37 | 2.17 | 8.10 | 2.73 |
| 21 | 36P | 2.2 | 0.3 | b.d | 4.17 | 0.262 | 9.91 | 3.08 | 8.53 | 2.95 |
| 22 | 37P | 1.2 | 0.1 | a | 3.44 | 0.541 | 8.96 | 1.58 | 6.37 | 2.87 |
| 23 | 40P | 1.8 | 0.3 | b,d | 4.89 | 0.633 | 11.54 | 1.80 | 10.82 | 2.53 |
| 24 | 41P | 0.7 | 0.1 | d | 3.09 | 0.660 | 9.23 | 1.05 | 5.42 | 2.83 |
| 25 | 42P | 0.7 | 0.1 | d | 4.86 | 0.585 | 3.99 | 2.01 | 10.70 | 2.63 |
| 26 | 43P | 2.0 | 0.4 | b,c,d | 3.35 | 0.595 | 15.97 | 1.36 | 6.13 | 2.79 |
| 27 | 44P | 1.7 | 0.3 | b,c,d | 3.69 | 0.427 | 5.90 | 2.12 | 7.10 | 2.92 |
| 28 | 45P | 0.7 | 0.6 | b,c,d | 3.02 | 0.825 | 4.25 | 0.53 | 5.25 | 2.58 |
| 29 | 46P | 0.6 | 0.3 | b,c,d | 3.09 | 0.658 | 11.74 | 1.06 | 5.44 | 2.82 |
| 30 | 47P | 3.1 | 0.2 | a | 4.12 | 0.318 | 13.04 | 2.81 | 8.36 | 2.91 |
| 31 | 48P | 3.0 | 0.2 | a | 3.65 | 0.370 | 13.70 | 2.30 | 6.98 | 2.94 |
| 32 | 49P | 4.2 | 0.7 | b,c,d | 3.56 | 0.600 | 19.05 | 1.42 | 6.73 | 2.71 |
| 33 | 50P | 1.5 | 0.1 | a | 4.09 | 0.529 | 19.16 | 1.92 | 8.27 | 2.69 |
| 34 | 52P | 1.2 | 0.2 | b,d | 3.84 | 0.543 | 10.22 | 1.76 | 7.54 | 2.77 |
| 35 | 53P | 3.3 | 0.5 | b,c,d | 5.39 | 0.552 | 6.61 | 2.41 | 12.53 | 2.65 |
| 36 | 54P | 2.4 | 0.4 | b | 3.79 | 0.427 | 6.07 | 2.17 | 7.38 | 2.91 |
| 37 | 56P | 1.9 | 0.1 | a | 5.11 | 0.504 | 8.16 | 2.53 | 11.54 | 2.71 |
| 38 | 57P | 1.0 | 0.1 | a | 3.45 | 0.500 | 2.85 | 1.72 | 6.40 | 2.92 |
| 39 | 58P | 0.6 | 0.1 | d | 4.08 | 0.662 | 13.48 | 1.38 | 8.24 | 2.57 |
| 40 | 59P | 0.8 | 0.1 | b,c,d | 4.49 | 0.475 | 9.34 | 2.36 | 9.51 | 2.77 |
| 41 | 60P | 0.7 | 0.1 | d | 3.51 | 0.538 | 3.61 | 1.62 | 6.56 | 2.86 |
| 42 | 61P | 0.9 | 0.4 | b,c,d | 3.68 | 0.427 | 6.01 | 2.11 | 7.06 | 2.93 |
| 43 | 62P | 0.6 | 0.1 | a | 3.53 | 0.578 | 10.50 | 1.49 | 6.63 | 2.80 |
| 44 | 63P | 1.4 | 0.2 | b,c,d | 5.58 | 0.651 | 19.78 | 1.95 | 13.19 | 2.41 |
| 45 | 64P | 1.7 | 0.3 | b,d | 4.44 | 0.690 | 8.95 | 1.38 | 9.35 | 2.49 |
| 46 | 65P | 4.6 | 2.0 | d | 3.75 | 0.301 | 10.30 | 2.63 | 7.27 | 2.98 |
| 47 | 67P | 2.1 | 0.3 | b,d | 3.46 | 0.641 | 7.04 | 1.24 | 6.45 | 2.74 |
| 48 | 68P | 2.8 | 0.2 | a | 4.90 | 0.641 | 11.14 | 1.76 | 10.83 | 2.52 |
| 49 | 69P | 0.9 | 0.1 | a | 3.88 | 0.414 | 22.05 | 2.27 | 7.65 | 2.80 |
| 50 | 70P | 1.7 | 0.4 | b,d | 3.68 | 0.453 | 6.60 | 2.01 | 7.06 | 2.90 |
| 51 | 71P | 0.9 | 0.1 | b,c,d | 3.13 | 0.498 | 9.48 | 1.57 | 5.53 | 2.99 |
| 52 | 73P | 1.1 | 0.3 | b,d | 3.06 | 0.695 | 11.42 | 0.93 | 5.34 | 2.78 |
| 53 | 74P | 3.3 | 0.7 | a | 4.16 | 0.148 | 6.65 | 3.55 | 8.50 | 3.01 |
| 54 | 75P | 1.7 | 0.1 | b,d | 3.54 | 0.496 | 5.91 | 1.78 | 6.67 | 2.89 |

| | | | | | | | | | |
|---|---|---|---|---|---|---|---|---|---|
| 55 | 76P | 0.3 | 0.5 | b | 3.47 | 0.539 | 30.48 | 1.60 | 6.47 | 2.68 |
| 56 | 77P | 1.7 | 0.1 | a | 3.60 | 0.358 | 24.40 | 2.31 | 6.83 | 2.86 |
| 57 | 78P | 1.4 | 0.1 | a | 3.74 | 0.462 | 6.26 | 2.01 | 7.22 | 2.89 |
| 58 | 79P | 0.7 | 0.1 | a | 2.95 | 0.619 | 3.15 | 1.12 | 5.06 | 2.95 |
| 59 | 81P | 2.3 | 0.1 | e | 3.45 | 0.537 | 3.24 | 1.60 | 6.41 | 2.88 |
| 60 | 82P | 0.7 | 0.1 | b,d | 4.14 | 0.123 | 1.13 | 3.63 | 8.43 | 3.03 |
| 61 | 84P | 1.0 | 0.2 | b,c,d | 3.65 | 0.492 | 7.28 | 1.85 | 6.97 | 2.87 |
| 62 | 86P | 0.7 | 0.1 | b,c,d | 3.63 | 0.366 | 15.45 | 2.30 | 6.91 | 2.93 |
| 63 | 87P | 0.4 | 0.2 | b,d | 3.49 | 0.376 | 2.58 | 2.17 | 6.51 | 3.01 |
| 64 | 88P | 1.0 | 0.1 | d | 3.11 | 0.562 | 4.38 | 1.36 | 5.48 | 2.95 |
| 65 | 89P | 1.4 | 0.1 | a | 3.80 | 0.399 | 12.03 | 2.28 | 7.39 | 2.90 |
| 66 | 90P | 2.6 | 0.4 | d | 6.04 | 0.509 | 9.62 | 2.97 | 14.85 | 2.69 |
| 67 | 91P | 1.3 | 0.2 | d | 3.90 | 0.329 | 14.08 | 2.62 | 7.70 | 2.92 |
| 68 | 92P | 1.4 | 0.3 | b,c,d | 5.37 | 0.663 | 18.76 | 1.81 | 12.43 | 2.41 |
| 69 | 93P | 2.6 | 0.3 | a | 4.39 | 0.612 | 12.22 | 1.70 | 9.20 | 2.61 |
| 70 | 94P | 2.3 | 0.2 | a | 3.51 | 0.364 | 6.18 | 2.24 | 6.58 | 3.00 |
| 71 | 97P | 2.0 | 0.5 | b,c,d | 4.80 | 0.458 | 17.88 | 2.60 | 10.51 | 2.71 |
| 72 | 98P | 1.0 | 0.1 | b | 3.81 | 0.561 | 10.54 | 1.67 | 7.43 | 2.76 |
| 73 | 99P | 4.8 | 0.8 | d | 6.15 | 0.229 | 4.33 | 4.74 | 15.25 | 2.96 |
| 74 | 100P | 2.5 | 0.4 | b | 3.42 | 0.417 | 25.66 | 1.99 | 6.31 | 2.85 |
| 75 | 101P | 1.0 | 0.1 | a | 5.79 | 0.594 | 5.08 | 2.35 | 13.94 | 2.59 |
| 76 | 103P | 0.6 | 0.0 | f | 3.47 | 0.695 | 13.62 | 1.06 | 6.47 | 2.64 |
| 77 | 104P | 1.2 | 0.5 | b,c,d | 3.37 | 0.585 | 15.49 | 1.40 | 6.18 | 2.80 |
| 78 | 105P | 0.8 | 0.1 | d | 3.47 | 0.409 | 9.17 | 2.05 | 6.46 | 2.97 |
| 79 | 106P | 0.9 | 0.2 | b,c,d | 3.77 | 0.587 | 20.11 | 1.56 | 7.31 | 2.68 |
| 80 | 108P | 0.8 | 0.1 | d | 3.75 | 0.542 | 13.08 | 1.72 | 7.26 | 2.78 |
| 81 | 110P | 2.1 | 0.3 | b,d | 3.62 | 0.313 | 11.68 | 2.48 | 6.88 | 2.99 |
| 82 | 111P | 1.2 | 0.2 | b,d | 4.16 | 0.110 | 4.23 | 3.70 | 8.49 | 3.02 |
| 83 | 112P | 0.9 | 0.2 | b,c | 3.54 | 0.586 | 24.17 | 1.46 | 6.67 | 2.69 |
| 84 | 113P | 1.7 | 0.1 | a | 3.69 | 0.423 | 5.78 | 2.13 | 7.09 | 2.93 |
| 85 | 114P | 0.9 | 0.2 | b,d | 3.54 | 0.556 | 18.28 | 1.57 | 6.67 | 2.77 |
| 86 | 115P | 1.1 | 0.1 | b,c | 4.26 | 0.521 | 11.69 | 2.04 | 8.79 | 2.73 |
| 87 | 116P | 3.1 | 0.5 | b | 3.48 | 0.374 | 3.61 | 2.18 | 6.48 | 3.01 |
| 88 | 117P | 3.6 | 2.9 | d | 4.09 | 0.254 | 8.70 | 3.06 | 8.29 | 2.97 |
| 89 | 118P | 1.3 | 0.2 | a | 3.47 | 0.428 | 8.51 | 1.98 | 6.45 | 2.96 |
| 90 | 119P | 1.0 | 0.1 | a | 4.28 | 0.292 | 5.20 | 3.03 | 8.84 | 2.94 |
| 91 | 120P | 0.8 | 0.1 | b,c,da | 4.13 | 0.339 | 8.80 | 2.73 | 8.39 | 2.92 |
| 92 | 121P | 3.9 | 0.3 | a | 4.64 | 0.190 | 20.15 | 3.75 | 9.98 | 2.86 |
| 93 | 123P | 2.2 | 0.2 | a | 3.86 | 0.449 | 15.36 | 2.13 | 7.59 | 2.83 |
| 94 | 124P | 2.6 | 0.2 | a | 3.32 | 0.504 | 31.53 | 1.64 | 6.04 | 2.74 |
| 95 | 125P | 0.8 | 0.1 | d | 3.13 | 0.512 | 9.99 | 1.53 | 5.53 | 2.97 |
| 96 | 127P | 0.9 | 0.1 | a | 3.45 | 0.363 | 14.31 | 2.19 | 6.40 | 2.98 |
| 97 | 128P | 2.3 | 0.4 | b,d | 4.49 | 0.321 | 4.36 | 3.05 | 9.51 | 2.91 |
| 98 | 129P | 1.2 | 0.1 | a | 4.32 | 0.093 | 3.43 | 3.91 | 8.96 | 3.02 |
| 99 | 130P | 2.2 | 0.2 | a | 3.54 | 0.412 | 7.34 | 2.08 | 6.67 | 2.96 |
| 100 | 131P | 1.1 | 0.1 | a | 3.68 | 0.343 | 7.35 | 2.42 | 7.06 | 2.98 |
| 101 | 132P | 0.8 | 0.1 | a | 4.09 | 0.530 | 5.77 | 1.92 | 8.28 | 2.77 |
| 102 | 134P | 1.5 | 0.2 | d | 6.23 | 0.587 | 4.35 | 2.57 | 15.55 | 2.60 |
| 103 | 135P | 1.4 | 0.2 | d | 3.83 | 0.290 | 6.05 | 2.72 | 7.50 | 2.99 |
| 104 | 136P | 1.1 | 0.3 | b,c | 4.19 | 0.293 | 9.43 | 2.96 | 8.59 | 2.93 |
| 105 | 137P | 4.0 | 0.3 | a | 4.50 | 0.576 | 4.88 | 1.91 | 9.55 | 2.67 |
| 106 | 138P | 0.8 | 0.1 | a | 3.62 | 0.531 | 10.09 | 1.70 | 6.89 | 2.83 |
| 107 | 139P | 1.4 | 0.1 | a | 4.52 | 0.247 | 2.33 | 3.40 | 9.61 | 2.96 |
| 108 | 143P | 4.8 | 0.3 | a | 4.30 | 0.410 | 4.69 | 2.54 | 8.93 | 2.86 |
| 109 | 144P | 0.8 | 0.1 | a | 3.87 | 0.628 | 4.11 | 1.44 | 7.60 | 2.68 |
| 110 | 146P | 1.0 | 0.1 | a | 4.03 | 0.648 | 23.08 | 1.42 | 8.08 | 2.53 |
| 111 | 147P | 0.2 | 0.1 | b | 3.81 | 0.276 | 2.37 | 2.76 | 7.43 | 3.01 |
| 112 | 148P | 1.1 | 0.1 | a | 3.68 | 0.538 | 3.68 | 1.70 | 7.07 | 2.83 |
| 113 | 149P | 1.4 | 0.1 | a | 4.33 | 0.388 | 29.73 | 2.65 | 9.02 | 2.66 |
| 114 | 152P | 2.1 | 0.3 | a | 4.50 | 0.307 | 9.87 | 3.12 | 9.54 | 2.90 |
| 115 | 154P | 1.7 | 0.2 | d | 4.84 | 0.672 | 18.06 | 1.59 | 10.66 | 2.43 |
| 116 | 159P | 1.5 | 0.2 | a | 5.90 | 0.381 | 23.42 | 3.65 | 14.32 | 2.69 |
| 117 | 160P | 1.0 | 0.1 | a | 3.97 | 0.479 | 17.28 | 2.07 | 7.92 | 2.77 |
| 118 | 162P | 7.0 | 0.5 | a | 3.05 | 0.597 | 27.84 | 1.23 | 5.32 | 2.79 |
| 119 | 163P | 1.3 | 0.8 | a | 3.76 | 0.453 | 12.72 | 2.06 | 7.30 | 2.86 |
| 120 | 168P | 0.5 | 0.1 | a | 3.62 | 0.610 | 21.93 | 1.41 | 6.90 | 2.66 |
| 121 | 169P | 2.5 | 0.1 | a | 2.60 | 0.767 | 11.31 | 0.61 | 4.20 | 2.89 |
| 122 | 171P | 1.3 | 0.1 | a | 3.55 | 0.503 | 21.95 | 1.76 | 6.69 | 2.79 |
| 123 | 172P | 5.7 | 0.4 | a | 3.51 | 0.362 | 1.53 | 2.24 | 6.58 | 3.01 |
| 124 | 173P | 4.3 | 1.2 | a | 5.71 | 0.262 | 16.49 | 4.21 | 13.63 | 2.85 |
| 125 | 179P | 2.4 | 0.4 | b | 5.89 | 0.307 | 19.89 | 4.08 | 14.30 | 2.79 |
| 126 | 184P | 0.6 | 0.1 | b | 3.57 | 0.594 | 1.53 | 1.45 | 6.75 | 2.79 |
| 127 | 197P | 0.9 | 0.1 | a | 2.87 | 0.630 | 25.54 | 1.06 | 4.85 | 2.86 |
| 128 | 213P | 1.5 | 0.3 | a | 3.42 | 0.380 | 10.24 | 2.12 | 6.33 | 3.00 |
| 129 | 215P | 1.3 | 0.1 | a | 4.02 | 0.201 | 12.79 | 3.22 | 8.07 | 2.97 |
| 130 | 216P | 0.6 | 0.1 | a | 3.88 | 0.445 | 9.04 | 2.15 | 7.65 | 2.87 |
| 131 | 219P | 1.0 | 0.1 | a | 3.65 | 0.353 | 11.52 | 2.36 | 6.98 | 2.96 |
| 132 | 221P | 1.0 | 0.1 | a | 3.48 | 0.486 | 11.41 | 1.79 | 6.50 | 2.90 |
| 133 | 223P | 2.9 | 0.2 | a | 4.14 | 0.418 | 27.05 | 2.41 | 8.42 | 2.70 |
| 134 | 228P | 1.2 | 0.2 | a | 4.16 | 0.176 | 7.92 | 3.43 | 8.49 | 2.99 |
| 135 | 243P | 1.5 | 0.2 | b | 3.83 | 0.359 | 7.64 | 2.46 | 7.50 | 2.95 |
| 136 | 246P | 4.2 | 0.4 | a | 4.02 | 0.287 | 15.99 | 2.87 | 8.06 | 2.91 |
| 137 | 264P | 11.0 | 2.3 | b | 3.89 | 0.374 | 25.15 | 2.44 | 7.68 | 2.79 |
| 138 | 256P | 0.8 | 0.1 | a | 4.62 | 0.419 | 27.66 | 2.68 | 9.92 | 2.64 |
| 139 | 260P | 1.5 | 0.1 | a | 3.69 | 0.594 | 15.74 | 1.50 | 7.07 | 2.72 |
| 140 | 269P | 5.0 | 0.9 | d | 6.75 | 0.436 | 6.63 | 4.05 | 17.52 | 2.81 |
| 141 | 270P | 2.5 | 0.4 | d | 7.22 | 0.467 | 2.86 | 3.60 | 19.41 | 2.80 |
| 142 | 280P | 2.5 | 0.3 | b | 4.51 | 0.418 | 11.78 | 2.62 | 9.58 | 2.81 |
| 143 | P/1997 G1 | 2.6 | 0.4 | d | 7.22 | 0.417 | 3.99 | 4.21 | 19.41 | 2.86 |
| 144 | P/2001 H5 | 1.0 | 0.1 | b | 5.99 | 0.600 | 8.40 | 2.40 | 14.68 | 2.57 |
| 145 | P/2001 R6 | 0.7 | 0.1 | a | 4.11 | 0.486 | 17.34 | 2.12 | 8.35 | 2.75 |
| 146 | P/2003 O3 | 0.6 | 0.1 | a | 3.10 | 0.599 | 8.36 | 1.25 | 5.47 | 2.90 |
| 147 | P/2003 S1 | 2.2 | 0.4 | b | 4.56 | 0.431 | 5.94 | 2.60 | 9.73 | 2.82 |
| 148 | P/2004 A1 | 3.5 | 0.3 | a | 7.90 | 0.308 | 10.58 | 5.46 | 22.20 | 2.96 |
| 149 | P/2004 D029 | 1.1 | 0.1 | a | 7.47 | 0.450 | 14.50 | 4.10 | 20.43 | 2.77 |
| 150 | P/2004 V5-A | 1.7 | 0.2 | a | 7.95 | 0.445 | 19.36 | 4.41 | 22.42 | 2.74 |
| 151 | P/2004 VR8 | 5.4 | 0.3 | a | 4.85 | 0.510 | 20.12 | 2.38 | 10.67 | 2.63 |
| 152 | P/2005 GF8 | 2.7 | 0.2 | a | 5.86 | 0.517 | 1.19 | 2.83 | 14.19 | 2.70 |
| 153 | P/2005 JD108 | 2.8 | 0.2 | a | 6.44 | 0.375 | 3.28 | 4.03 | 16.36 | 2.87 |
| 154 | P/2005 L4 | 1.9 | 0.1 | a | 4.11 | 0.425 | 17.04 | 2.37 | 8.35 | 2.80 |
| 155 | P/2005 Q4 | 1.5 | 0.1 | a | 4.46 | 0.607 | 17.65 | 1.75 | 9.42 | 2.57 |
| 156 | P/2005 R1 | 1.7 | 0.1 | a | 5.50 | 0.628 | 15.40 | 2.05 | 12.91 | 2.49 |
| 157 | P/2005 S3 | 1.4 | 0.1 | a | 4.90 | 0.420 | 3.48 | 2.84 | 10.86 | 2.82 |
| 158 | P/2005 T5 | 1.3 | 0.1 | a | 7.25 | 0.552 | 21.37 | 3.25 | 19.54 | 2.55 |
| 159 | P/2005 W3 | 1.2 | 0.3 | a | 6.41 | 0.530 | 16.78 | 3.01 | 16.22 | 2.61 |
| 160 | P/2005 Y2 | 5.1 | 0.4 | a | 6.30 | 0.467 | 19.18 | 3.36 | 15.81 | 2.66 |
| 161 | D/1819 W1 | 0.2 | 0.4 | b | 2.96 | 0.699 | 9.11 | 0.89 | 5.10 | 2.82 |

# TABLE 2

Table 2. Size-frequency distributions and uncertainties

| Effective radius (km) | $J_{obs}(r)$ | Uncertainty in $J_{obs}(r)$ | Observational Correction factor (OCF) | $J(r)$ | $SD(r)$ (A'=500, B'=6) |
|---|---|---|---|---|---|
| 0.2 | 8.66 | 6.58 | 22.40 | 193.97 | 3006608 |
| 0.4 | 13.08 | 8.09 | 16.26 | 212.72 | 1701722 |
| 0.6 | 75.11 | 19.38 | 11.45 | 860.08 | 4730454 |
| 0.8 | 94.47 | 21.73 | 7.78 | 734.98 | 3123671 |
| 1.0 | 94.87 | 21.78 | 5.69 | 540.17 | 1890605 |
| 1.2 | 67.20 | 18.33 | 4.62 | 310.55 | 931653 |
| 1.4 | 68.14 | 18.46 | 3.96 | 270.01 | 713603 |
| 1.6 | 52.73 | 16.24 | 3.50 | 184.54 | 438289 |
| 1.8 | 43.66 | 14.77 | 3.31 | 144.72 | 313556 |
| 2.0 | 33.12 | 12.87 | 3.13 | 103.76 | 207518 |
| 2.2 | 33.00 | 12.84 | 3.06 | 100.94 | 188115 |
| 2.4 | 35.21 | 13.27 | 2.99 | 105.12 | 183958 |
| 2.6 | 31.23 | 12.50 | 2.90 | 90.53 | 149721 |
| 2.8 | 28.87 | 12.01 | 2.79 | 80.66 | 126746 |
| 3.0 | 19.61 | 9.90 | 2.69 | 52.75 | 79129 |
| 3.2 | 11.73 | 7.66 | 2.57 | 30.20 | 43412 |
| 3.4 | 7.35 | 6.06 | 2.42 | 17.75 | 24536 |
| 3.6 | 6.64 | 5.76 | 2.26 | 14.99 | 19987 |
| 3.8 | 7.00 | 5.92 | 2.10 | 14.69 | 18941 |
| 4.0 | 6.91 | 5.88 | 1.94 | 13.44 | 16799 |
| 4.2 | 6.13 | 5.54 | 1.85 | 11.33 | 13756 |
| 4.4 | 5.33 | 5.16 | 1.75 | 9.34 | 11033 |
| 4.6 | 5.07 | 5.04 | 1.66 | 8.40 | 9675 |
| 4.8 | 5.11 | 5.06 | 1.56 | 7.97 | 8969 |
| 5.0 | 5.06 | 5.03 | 1.46 | 7.41 | 8150 |
| 5.2 | 4.92 | 4.96 | 1.40 | 6.88 | 7412 |
| 5.4 | 4.58 | 4.78 | 1.34 | 6.12 | 6465 |
| 5.6 | 3.82 | 4.37 | 1.28 | 4.87 | 5048 |
| 5.8 | 2.81 | 3.75 | 1.23 | 3.45 | 3505 |
| 6.0 | 1.94 | 3.12 | 1.18 | 2.29 | 2291 |
| 6.2 | 1.39 | 2.64 | 1.13 | 1.57 | 1548 |
| 6.4 | 1.13 | 2.37 | 1.09 | 1.23 | 1193 |
| 6.6 | 1.07 | 2.31 | 1.07 | 1.14 | 1092 |
| 6.8 | 1.09 | 2.33 | 1.05 | 1.14 | 1076 |
| 7.0 | 1.08 | 2.32 | 1.03 | 1.11 | 1028 |
| 7.2 | 0.97 | 2.21 | 1.01 | 0.98 | 898 |
| 7.4 | 0.79 | 1.99 | 1.00 | 0.79 | 714 |
| 7.6 | 0.58 | 1.71 | 1.00 | 0.58 | 520 |
| 7.8 | 0.40 | 1.42 | 1.00 | 0.40 | 357 |
| 8.0 | 0.28 | 1.19 | 1.00 | 0.28 | 248 |
| 8.2 | 0.22 | 1.04 | 1.00 | 0.22 | 188 |
| 8.4 | 0.19 | 0.97 | 1.00 | 0.19 | 160 |
| 8.6 | 0.18 | 0.94 | 1.00 | 0.18 | 149 |
| 8.8 | 0.17 | 0.93 | 1.00 | 0.17 | 147 |
| 9.0 | 0.18 | 0.94 | 1.00 | 0.18 | 147 |
| 9.2 | 0.18 | 0.95 | 1.00 | 0.18 | 149 |
| 9.4 | 0.19 | 0.97 | 1.00 | 0.19 | 152 |
| 9.6 | 0.19 | 0.98 | 1.00 | 0.19 | 156 |
| 9.8 | 0.20 | 1.00 | 1.00 | 0.20 | 159 |
| 10.0 | 0.20 | 1.01 | 1.00 | 0.20 | 163 |
| 10.2 | 0.21 | 1.02 | 1.00 | 0.21 | 166 |
| 10.4 | 0.22 | 1.04 | 1.00 | 0.22 | 169 |
| 10.6 | 0.22 | 1.05 | 1.00 | 0.22 | 172 |
| 10.8 | 0.22 | 1.06 | 1.00 | 0.22 | 174 |
| 11.0 | 0.23 | 1.07 | 1.00 | 0.23 | 175 |
| 11.2 | 0.23 | 1.07 | 1.00 | 0.23 | 176 |
| 11.4 | 0.23 | 1.08 | 1.00 | 0.23 | 176 |
| 11.6 | 0.23 | 1.08 | 1.00 | 0.23 | 175 |
| 11.8 | 0.23 | 1.08 | 1.00 | 0.23 | 174 |
| 12.0 | 0.23 | 1.07 | 1.00 | 0.23 | 172 |
| 12.2 | 0.23 | 1.07 | 1.00 | 0.23 | 170 |
| 12.4 | 0.23 | 1.06 | 1.00 | 0.23 | 167 |
| 12.6 | 0.22 | 1.05 | 1.00 | 0.22 | 163 |
| 12.8 | 0.22 | 1.04 | 1.00 | 0.22 | 160 |
| 13.0 | 0.21 | 1.03 | 1.00 | 0.21 | 156 |
| 13.2 | 0.21 | 1.02 | 1.00 | 0.21 | 151 |
| 13.4 | 0.20 | 1.01 | 1.00 | 0.20 | 147 |
| 13.6 | 0.20 | 1.00 | 1.00 | 0.20 | 143 |
| 13.8 | 0.19 | 0.98 | 1.00 | 0.19 | 138 |
| 14.0 | 0.19 | 0.97 | 1.00 | 0.19 | 134 |
| 14.2 | 0.18 | 0.96 | 1.00 | 0.18 | 130 |
| 14.4 | 0.18 | 0.94 | 1.00 | 0.18 | 126 |
| 14.6 | 0.17 | 0.93 | 1.00 | 0.17 | 122 |
| 14.8 | 0.17 | 0.92 | 1.00 | 0.17 | 118 |
| 15.0 | 0.16 | 0.90 | 1.00 | 0.16 | 114 |
| 15.2 | 0.16 | 0.89 | 1.00 | 0.16 | 110 |
| 15.4 | 0.15 | 0.88 | 1.00 | 0.15 | 107 |

**Figure 1**

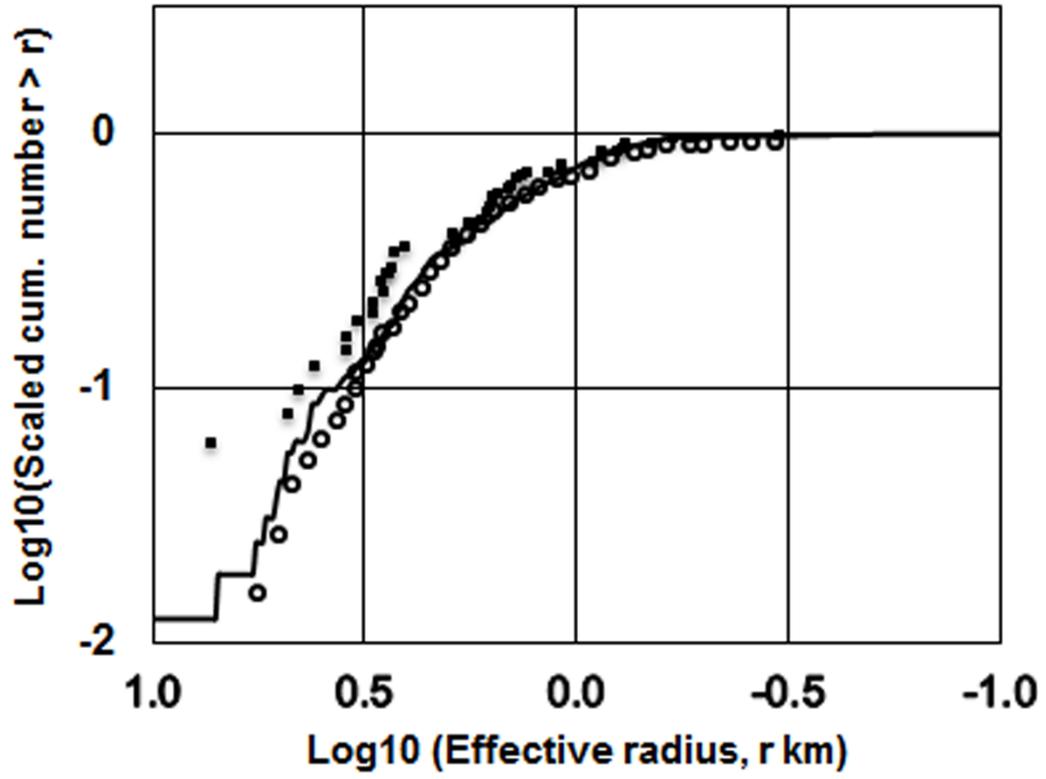

**Figure 2**

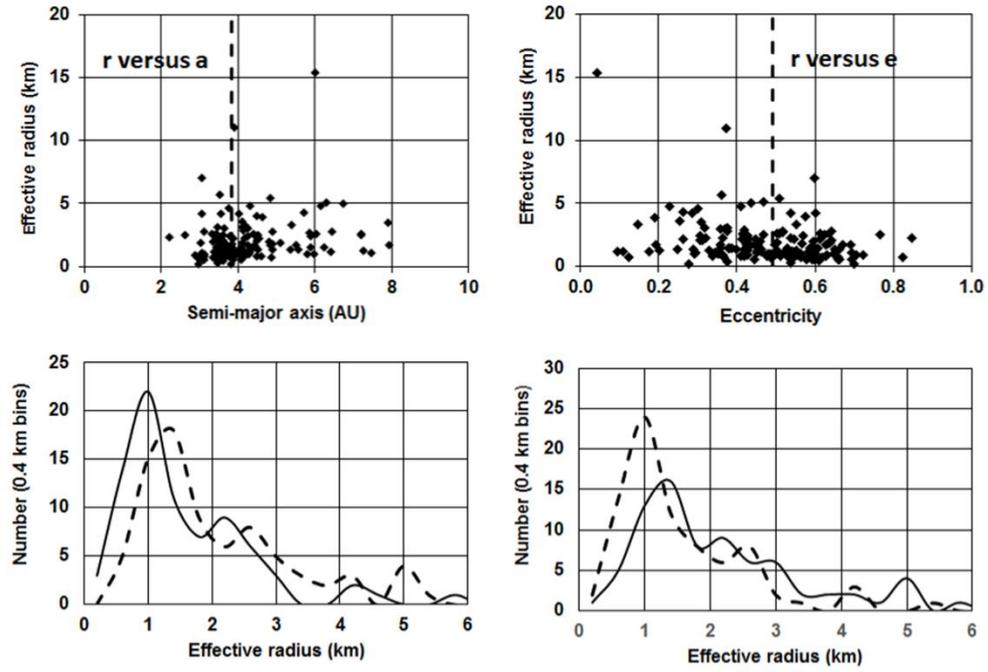

**Figure 3**

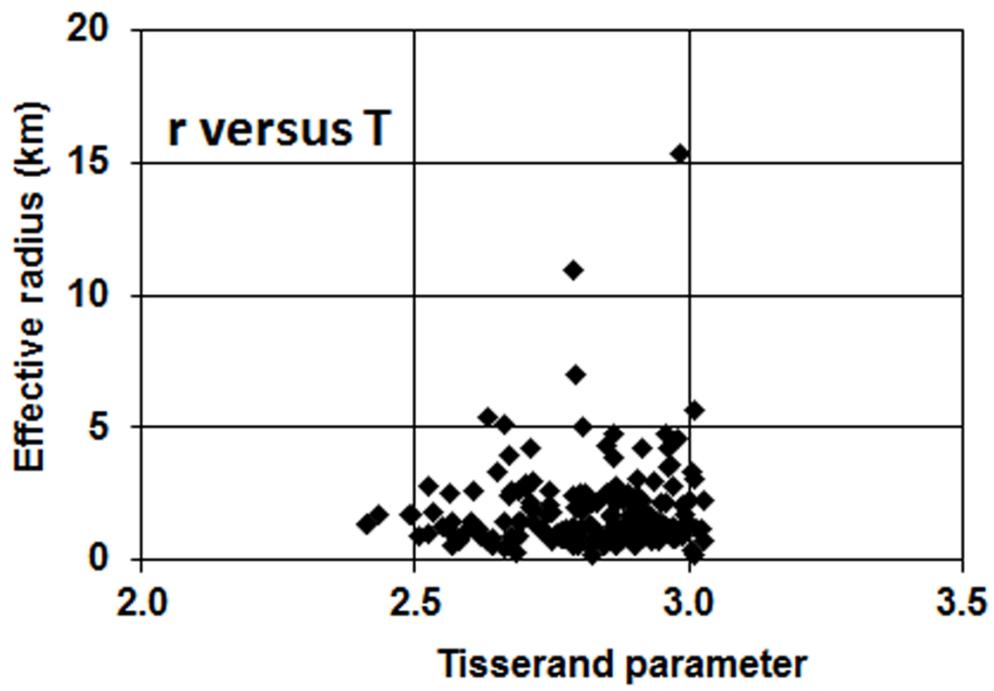

**Figure 4**

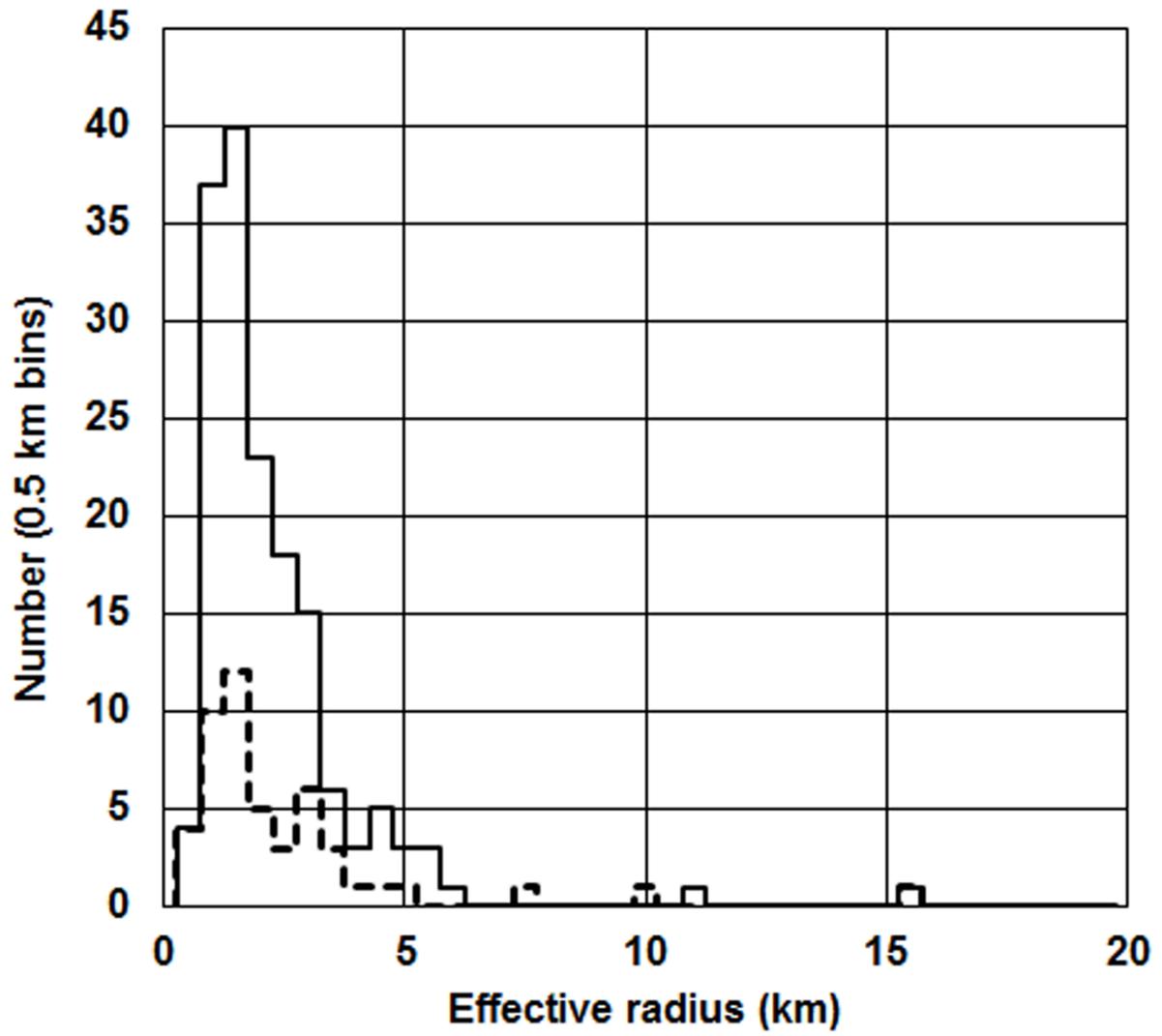

**Figure 5**

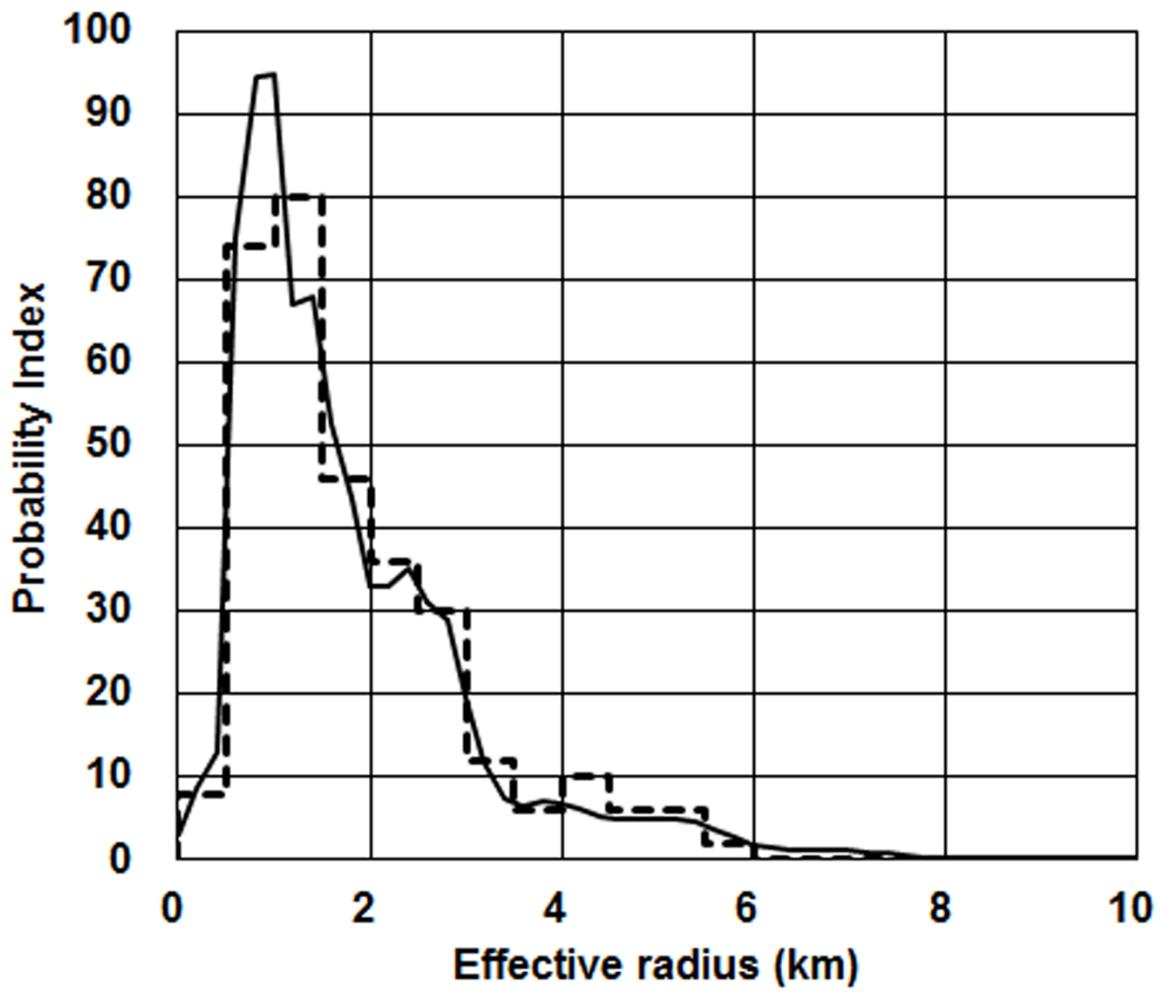

**Figure 6**

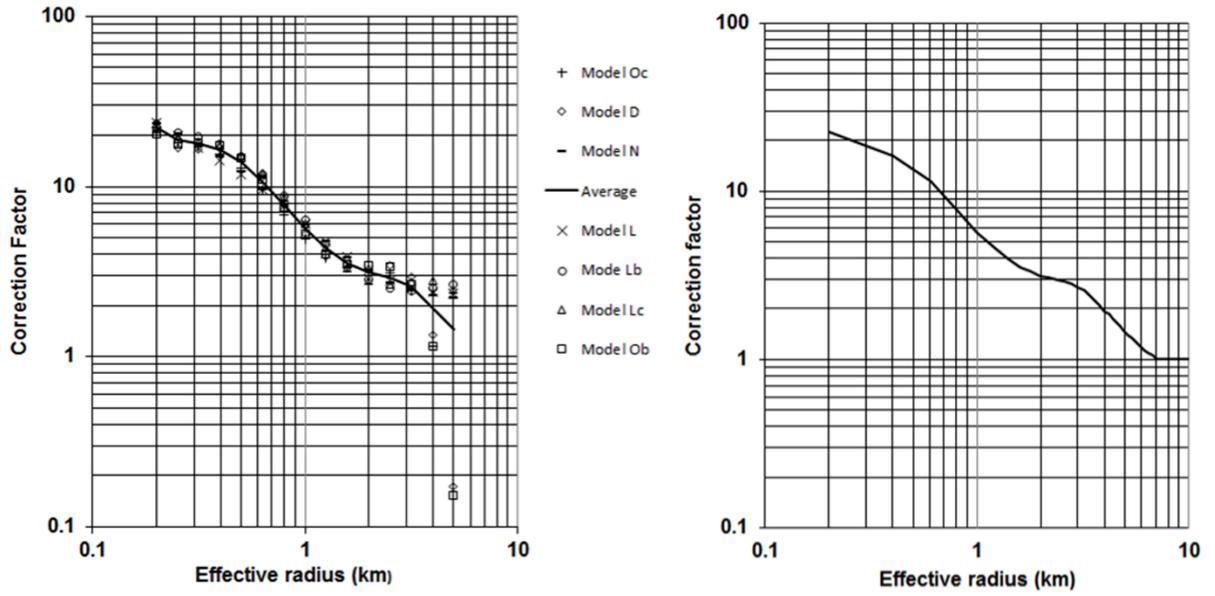

**Figure 7**

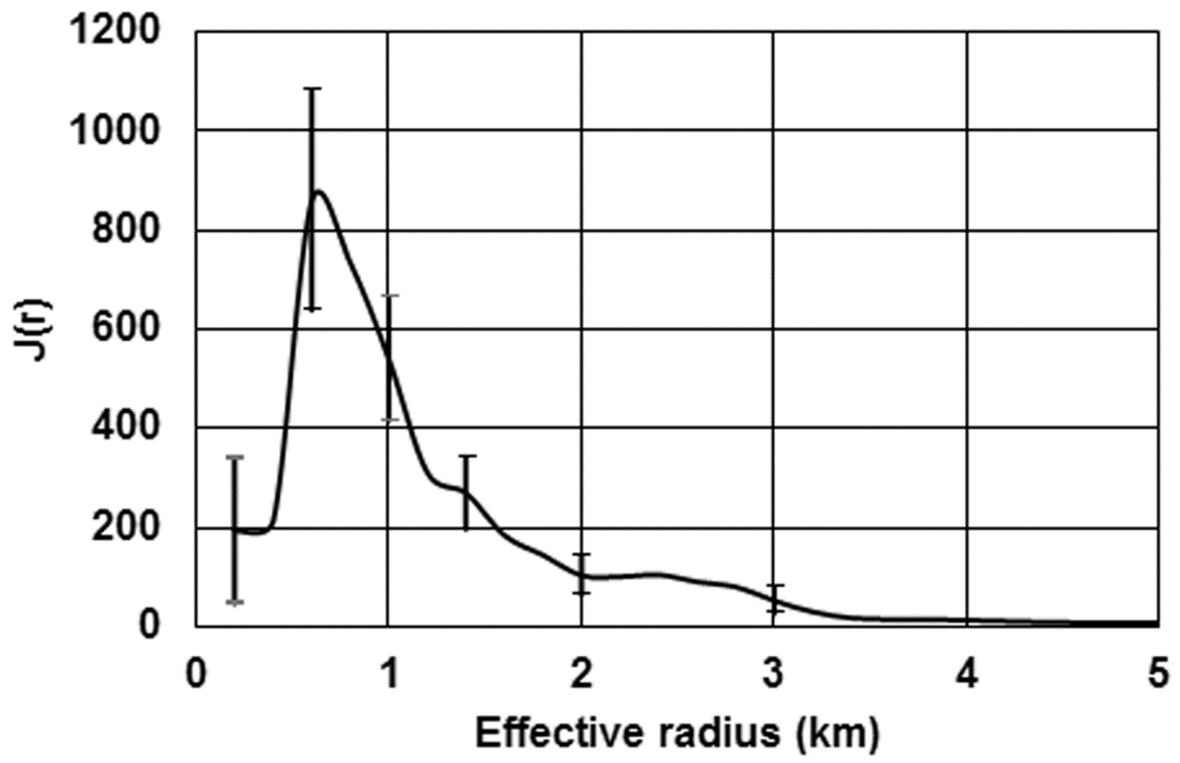

**Figure 8**

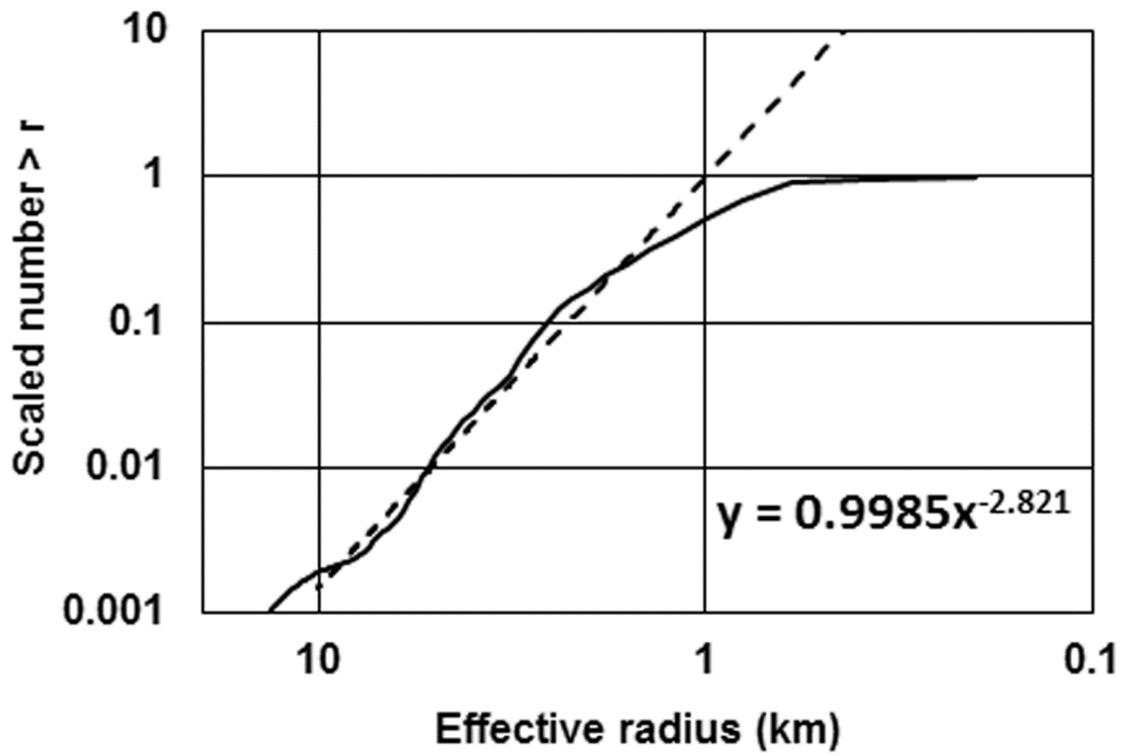

**Figure 9**

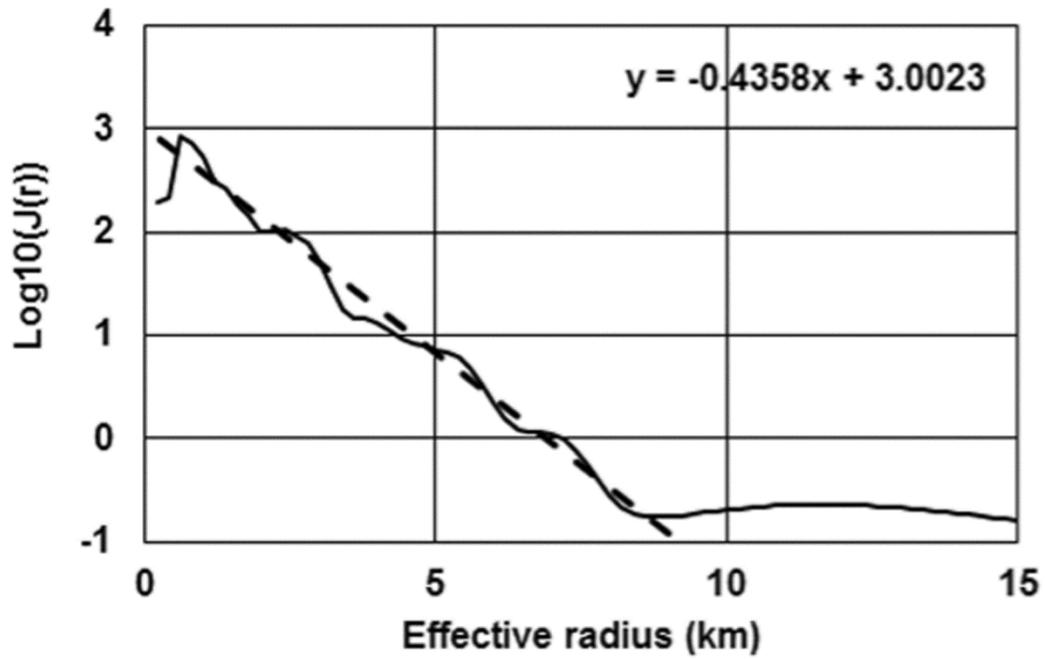

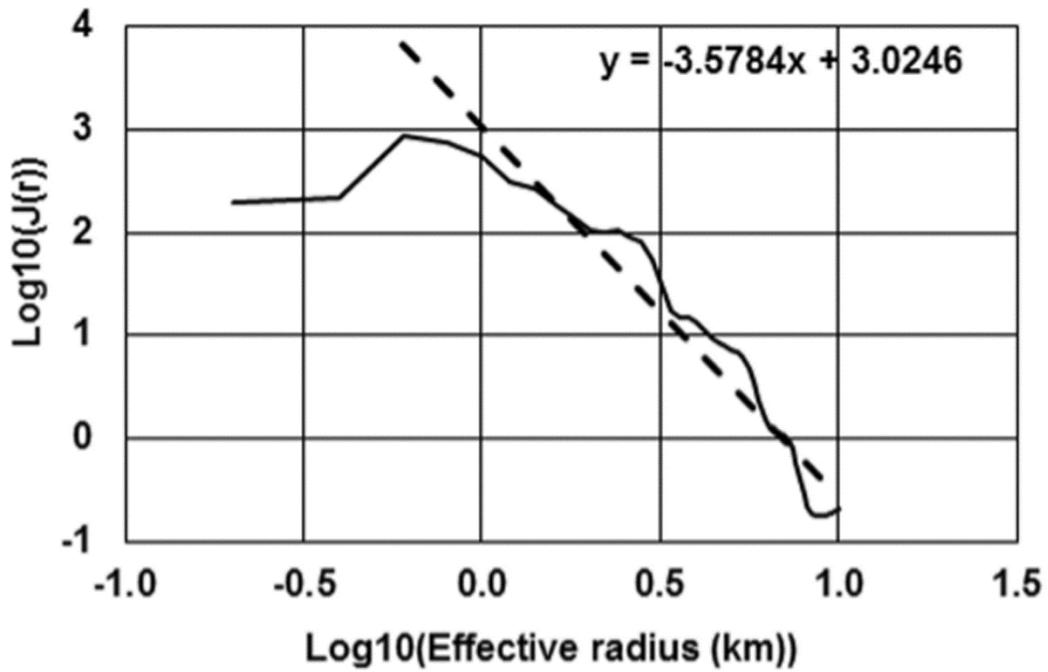

**Figure 10**

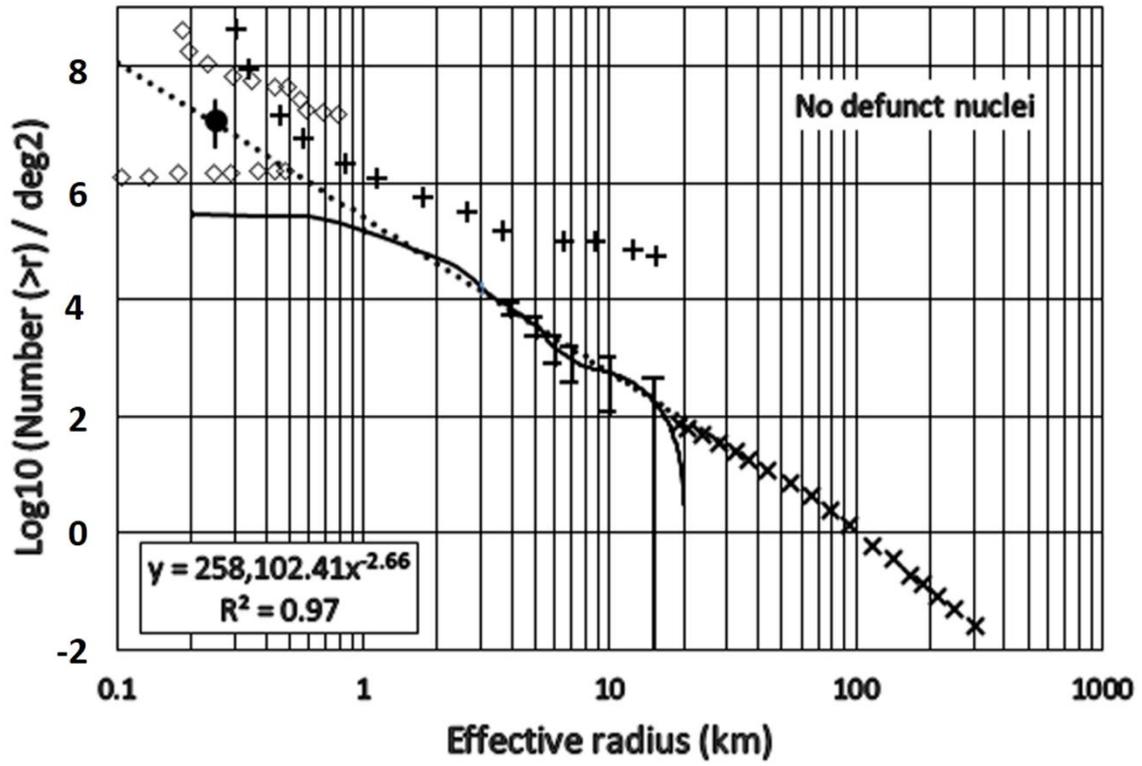

**Figure 11**

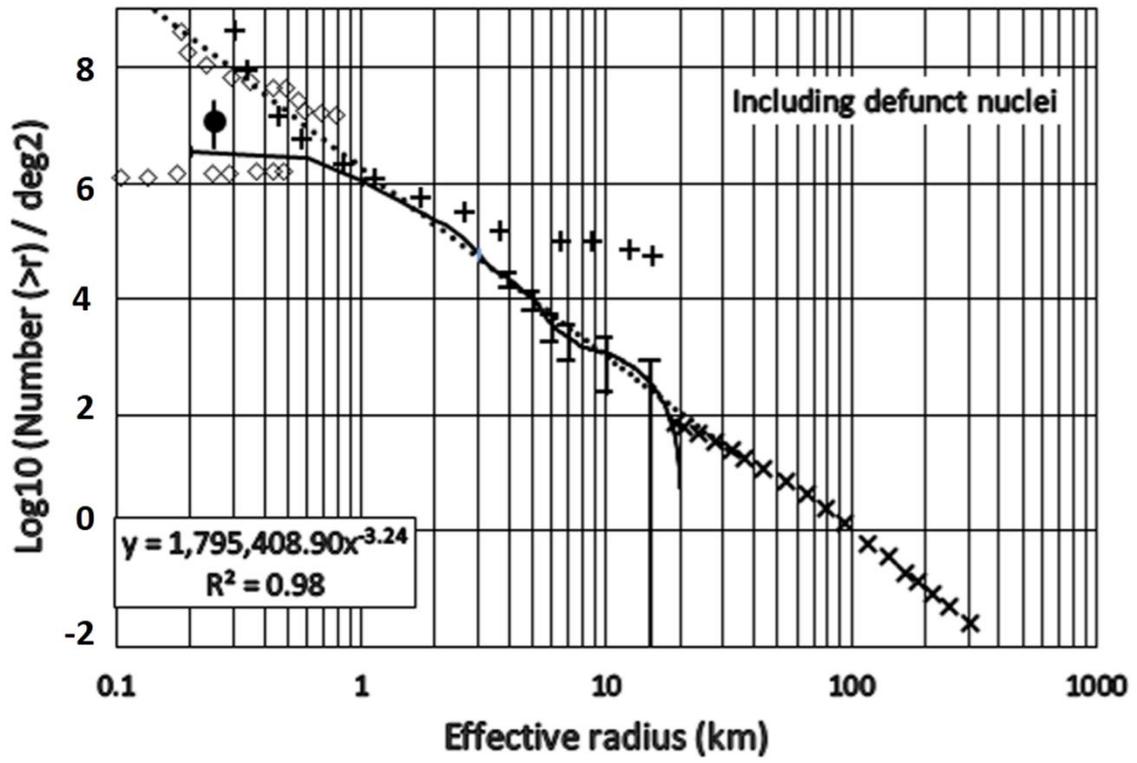

**Figure 12**

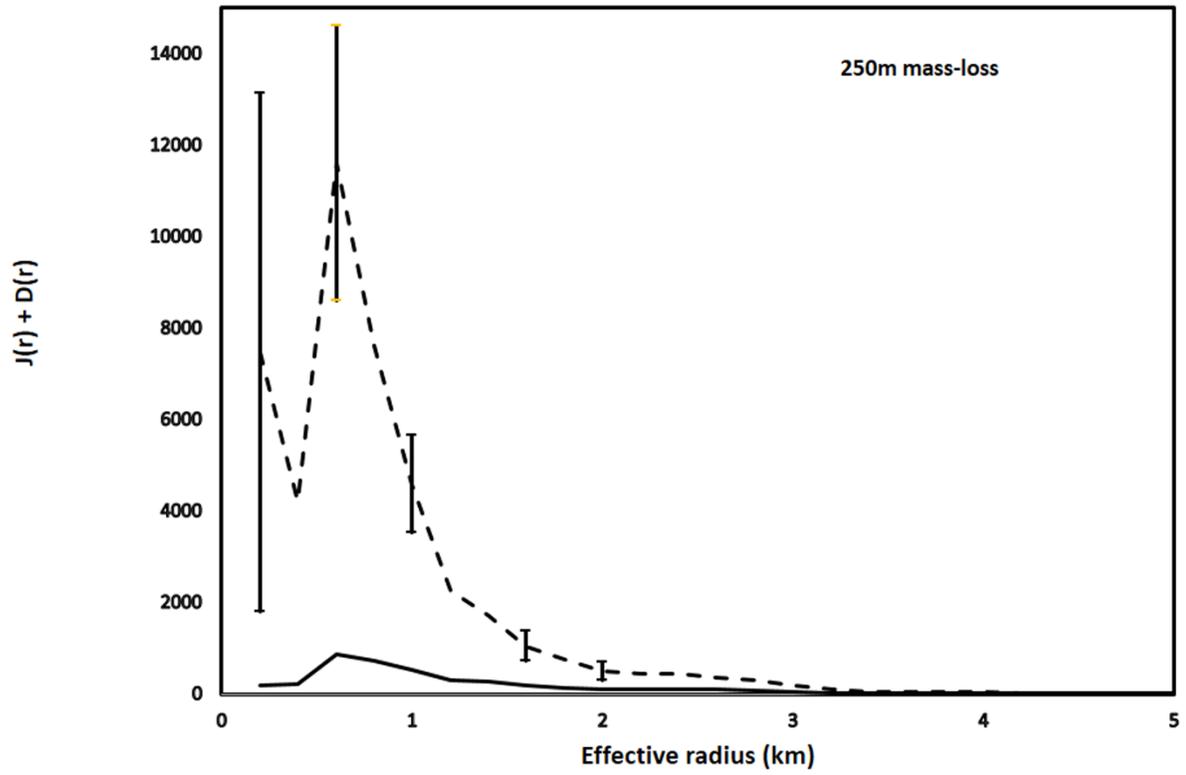